\documentclass[jkps,preprint,fleqn]{revtex4} 
\usepackage{graphicx}
\usepackage{amssymb}
\usepackage{amsmath}
\usepackage{bm}

\usepackage{kotex}

\pdfoutput=1

\usepackage{float} \usepackage{graphicx}
\usepackage{epstopdf}
\usepackage{graphics}
\usepackage{caption}
\usepackage{subfigure}
\usepackage{epsfig}
\usepackage{color}
\usepackage{multirow,array}

\usepackage{tensor}

\usepackage{adjustbox}  

\usepackage{tabularray}

\usepackage{hyperref}
\hypersetup{
	colorlinks=true,
	linkcolor=blue,
	filecolor=magneta,      
	urlcolor=blue,
}

\newcommand{\GR}{\textrm{GR}}

\newcommand{\be}{\begin{equation}}
\newcommand{\ee}{\end{equation}}
\newcommand{\bea}{\begin{eqnarray}}
\newcommand{\eea}{\end{eqnarray}}

\newcommand{\tc}{{\tilde c}}
\newcommand{\tG}{{\tilde G}}

\newcommand{\tv}{\textrm{V}}
\newcommand{\tT}{\textrm{T}}

\begin{document}
\setcounter{page}{0}
\title[]{Perturbation Theory in the minimally extended Varying Speed of Light (meVSL) Model }
\author{Seokcheon \surname{Lee}}
\email{skylee@skku.edu}
\affiliation{Department of Physics, Institute of Basic Science, Sungkyunkwan University, Suwon 16419, Korea}

\date[]{Received }

\begin{abstract}
Cosmological perturbation theory is a fundamental framework for analyzing the evolution of density fluctuations and gravitational potentials in the Universe. It plays a crucial role in understanding large-scale structure formation and the cosmic microwave background anisotropies. In this study, we apply perturbation theory to the minimally extended varying speed of light (meVSL) model to investigate the effects of a varying speed of light on the matter density contrast and the Newtonian gravitational potential. Unlike conventional models where the speed of light is constant, the meVSL model introduces modifications to the cosmological evolution equations, leading to potential deviations in structure formation and gravitational interactions. By deriving and analyzing the perturbed equations within this framework, we aim to explore how a varying speed of light influences the growth of density perturbations and the evolution of gravitational potentials. Our results provide new insights into the theoretical implications of the meVSL model and its possible observational signatures, particularly in resolving cosmological tensions such as the Hubble discrepancy.
\end{abstract}



\maketitle


\section{Introduction}\label{sec1}

Cosmological perturbation theory provides a powerful framework for studying the evolution of density fluctuations and gravitational potentials in the universe \cite{Lifshitz46,Zeldovich:1969sb,Stewart:1974uz,Bardeen:1980kt,Kodama:1984ziu,Mukhanov:1990me,Ma:1995ey,Bertschinger:2001is,Durrer:2004fx}. It is essential for understanding the formation of large-scale structures (LSS) \cite{Davis:1992ui,Bertschinger:1998tv}, the anisotropies in the cosmic microwave background (CMB) \cite{Sachs:1967er,Tomita:1982yd,Abbott:1986ct,Atrio-Barandela:1994ewh,White:1994sx,Kamionkowski:1996ks,Hu:1997mn}, and the growth of cosmic inhomogeneities \cite{Peebles:1970ag,Silk:1980cja,Baugh:1994hb,Bernardeau:1994aq}. Within the standard $\Lambda$CDM paradigm, the evolution of perturbations is governed by Einstein’s field equations, where the speed of light, $c$, is assumed to be constant. However, alternative models that allow for a varying speed of light (VSL) have been proposed as possible extensions to standard cosmology, offering novel mechanisms to address fundamental issues such as the horizon problem \cite{Petit:1988,Petit:1988-2,Petit:1989,Midy:1989,Moffat:1992ud,Petit:1995ass,Albrecht:1998ir,Barrow:1998he,Clayton:1998hv,Barrow:1999jq,Barrow:1999st,Clayton:1999zs,Brandenberger:1999bi,Bassett:2000wj,Clayton:2000xt,Gopakumar:2000kp,Jacobson:2000xp,Magueijo:2000zt,Magueijo:2000au,Magueijo:2003gj,Magueijo:2007gf,Petit:2008eb,Roshan:2009yb,Sanejouand:2009,Nassif:2012dr,Moffat:2014poa,Ravanpak:2017kdg,Costa:2017abc,Nassif:2018pdu}, some of the limitations of standard cosmology \cite{Avelino:2002jn,Pedram:2007mj,Salzano:2016pny,Leszczynska:2018juk}, and the nature of dark energy \cite{Avelino:1999is,Belinchon:1999kq,Drummond:1999ut,Alexander:1999cb,Avelino:2000ph,Szydlowski:2002kz,Shojaie:2004sq,Shojaie:2004xw,Balcerzak:2013kha,Balcerzak:2014rga,Franzmann:2017nsc,Hanimeli:2019wrt,Skara:2019usd,Bhattacharjee:2020fgl,Gupta:2020anq,Cuzinatto:2022mfe,Cuzinatto:2022vvy,Cuzinatto:2022dta}.   

The minimally extended varying speed of light (meVSL) model preserves the conventional Friedmann-Lema$\hat{\i}$tre-Robertson-Walker (FLRW) metric while introducing a general condition on cosmological time dilation (CTD), allowing the speed of light to vary with time. Instead of modifying the metric itself, the meVSL framework naturally leads to a time-dependent speed of light as a consequence of this condition, without requiring a specific functional form for $c(a)$. This distinguishes the meVSL model from traditional VSL models, which either propose or rely on an explicit mechanism for varying the speed of light \cite{Lee:2023bjz,Lee:2024mal}.

The meVSL model is built upon the FLRW metric, which satisfies the cosmological principle (CP) by ensuring spatial isotropy and homogeneity. Maintaining adiabaticity is essential to preserve these symmetries, as any net energy flux would introduce a preferential energy flow direction, violating isotropy \cite{Lee:2022heb}. Consequently, if the speed of light varies over cosmic time, the Planck constant must also evolve accordingly.  To ensure consistency with all known local physical laws, including special relativity and electromagnetism, cosmological evolution of other physical constants and quantities must be induced \cite{Lee:2024mal,Lee:2020zts}.

So far, discussions on the meVSL model have primarily focused on its implications at the background level in cosmological observations \cite{Lee:2023rqv,Lee:2023ucu,Lee:2024nya,Lee:2021ona,Lee:2024kxa}. However, despite its theoretical advantages, the effects of meVSL on structure formation and gravitational interactions remain largely unexplored \cite{Lee:2024zcu}. Since cosmological perturbations play a crucial role in shaping the universe, it is essential to investigate how a varying speed of light influences the evolution of matter density fluctuations and the Newtonian gravitational potential. In particular, modifications to the perturbation equations could lead to observable consequences, such as changes in the growth rate of large-scale structures, alterations in the CMB power spectrum, and potential deviations in gravitational lensing effects. Therefore, it is necessary to study linear perturbation theory within the meVSL framework as a prerequisite.

In this work, we systematically apply cosmological perturbation theory to the meVSL framework to derive the perturbed evolution equations for the matter density contrast and the gravitational potential. We analyze the influence of a time-dependent $c(a)$ on the growth of density perturbations and explore whether these modifications could provide new insights into open problems in cosmology. By comparing the predictions of the meVSL model with standard $\Lambda$CDM results, we aim to identify potential observational signatures that could distinguish this model from conventional cosmological scenarios.

The structure of this paper is as follows. In section $1$, we provide an introduction to the study. Section $2$ develops the linear cosmological perturbation theory within the meVSL framework, including perturbations of the FLRW metric, the energy-momentum tensor, coordinate and gauge transformations, and the scalar-vector-tensor decomposition. In section $3$, we derive Einstein field equations for background and scalar perturbations. Finally, we discuss the implications of our results and present our conclusions in Section $4$.

\section{Linear Perturbation Theory of the FLRW Metric in the meVSL Model} 
\label{sec:metricPT}

The CP, which assumes a homogeneous and isotropic universe, is a powerful and productive framework for understanding large-scale cosmic phenomena. However, this assumption is valid primarily on scales larger than $200$ Mpc. The universe deviates from perfect homogeneity and isotropy on smaller scales, such as those associated with galaxies and galaxy clusters. To study these deviations, we consider perturbations within the FLRW metric. Given the compelling observational evidence for a spatially flat universe, we simplify our analysis by adopting a flat geometry.

\subsection{Perturbations of the FLRW metric}
\label{subsec:perturbations}

The metric with perturbations is expressed as
\begin{align}
g_{\mu\nu} = \begin{pmatrix} -1 + h_{00} & a h_{0i} \\ a h_{i0} & \bar{g}_{ij} + a^2 h_{ij} \end{pmatrix} = \begin{pmatrix} -1 + h_{00} & a h_{0i} \\ a h_{i0} & a^2 \left( \gamma_{ij} + h_{ij} \right) \end{pmatrix} \equiv \bar{g}_{\mu\nu} + \delta g_{\mu\nu} \label{gmunuwhmunu} \,,
\end{align}
where \( \bar{g}_{\mu\nu} \) represents the background FLRW metric, with \( \gamma_{ij} = \delta_{ij} \) corresponding to a flat universe, and \( |\delta g_{\mu\nu}| \ll |\bar{g}_{\mu\nu} |\) is the perturbation term. To manipulate the perturbed metric, we use the following relations
\begin{align}
	g^{\mu\rho}g_{\rho\nu} = \delta^\mu{}_\nu \quad, \quad \bar{g}^{\mu\rho}\bar{g}_{\rho\nu} = \delta^\mu{}_\nu \quad, \quad
	\delta g^{\mu\nu} = - \bar{g}^{\mu\rho}\delta g_{\rho\sigma}\bar{g}^{\nu\sigma} \label{metricdifApp} \,.
\end{align}
These relations ensure consistency between the covariant and contravariant components of the metric. They describe how the perturbation \( \delta g_{\mu\nu} \) is related to the inverse background metric \( \bar{g}^{\mu\nu} \). Using these, we can express the perturbed contravariant metric as
\begin{equation}
	\delta g^{00} = - h_{00}\;, \quad \delta g^{0i} = - \bar{g}^{00}\delta g_{0j}\bar{g}^{ij} = \frac{1}{a}h_{0j} \gamma^{ij}\;, \quad \delta g^{ij} = - \bar{g}^{ik}\delta g_{kl}\bar{g}^{jl} = -\frac{1}{a^2} \gamma^{ik} h_{kl} \gamma^{jl} \;,
\end{equation}
where \( \gamma^{ij} = \delta^{ij} \) is the inverse spatial metric, and \( h_{ij} \) represents the spatial components of the perturbations. These equations show how the perturbation terms \( h_{00} \), \( h_{0i} \), and \( h_{ij} \) manifest in the contravariant components of the metric. We have assumed that the indices of \( h_{ij} \) are raised using the Kronecker delta \( \delta^{ij} \), and that \( h^{ij} = h_{ij} \). Finally, it is important to note that \( \delta g_{\mu\nu} \) refers to the perturbed covariant metric, while \( \delta g^{\mu\nu} \)  is not the contravariant perturbed metric.

We describe the spacetime in the background by the background metric, which is a theoretical construct. However, this background spacetime does not fully represent the physical reality, as it does not account for the deviations from homogeneity and isotropy. Instead, one can express these deviations between the actual physical spacetime and the background as
\begin{equation}\label{metricdecApp}
	\delta g_{\mu\nu}(x^{\sigma}) = g_{\mu\nu}(x^{\sigma}) - \bar{g}_{\mu\nu}(x^{\sigma}) \,,
\end{equation}
evaluated at a specific spacetime coordinate \( x^{\sigma} \). However, the concept of \( x^{\sigma} \) becomes problematic because \( g_{\mu\nu} \) and \( \bar{g}_{\mu\nu} \) are tensors defined on different manifolds (i.e., different spaces), and \( x^{\sigma} \) refers to coordinates defined in different charts (i.e., coordinate systems). Comparing tensors evaluated at different points is not well-defined, even when we embed both manifolds into a common one.  We need to establish a map that relates points in the background manifold to those in the physical manifold to make equation \eqref{metricdecApp} meaningful. This map called a gauge, allows us to describe points in the physical manifold using the same coordinate system as in the background manifold. As a result, cosmic time and comoving spatial coordinates can also be used to describe the perturbations. This brings up the issue known as the gauge problem, which we will briefly discuss later.

\subsubsection{The perturbed Christoffel symbols}
\label{subsubsec:Christoffel symbols}

The Christoffel symbols derived from the general metric in Eq.~\eqref{gmunuwhmunu} can be decomposed into the background and perturbed components as 
\begin{align}
	\Gamma^\mu_{\nu\rho} &= \frac{1}{2} g^{\mu\sigma} \left( g_{\sigma\rho,\nu} + g_{\nu\sigma,\rho} - g_{\nu\rho,\sigma} \right) \equiv \bar{\Gamma}^\mu_{\nu\rho} + \delta \Gamma^\mu_{\nu\rho} \quad , \textrm{where} \nonumber \\
	\delta \Gamma^\mu_{\nu\rho} &= \frac{1}{2} \bar{g}^{\mu\sigma} \left( \delta g_{\sigma\nu,\rho} + \delta g_{\sigma\rho,\nu} - \delta g_{\nu\rho,\sigma} - 2 \delta g_{\sigma\alpha} \bar{\Gamma}^\alpha_{\nu\rho} \right) \,, \label{Chirstwh}
\end{align}
where the background Christoffel symbols \( \bar{\Gamma}^\mu_{\nu\rho} \) are computed solely from the background metric \( \bar{g}_{\mu\nu} \).  In the $4$-vector representation, the time component is given by \( x^0 = c t \). In the meVSL model, the speed of light \( c \) is a function of time. If we take the differential of \( x^0 \), we obtain \cite{Lee:2024zcu}
\begin{align}
dx^0 = d \left(c t \right) = \left( \frac{d c}{dt} \frac{t}{c} + 1 \right) c dt \equiv N c dt \equiv \tilde{c} dt \,. \label{dx01}
\end{align}
Therefore, when differentiating physical quantities, including the Christoffel symbols, with respect to \( x^0 \), we must take into account the following relation
\begin{align}
\frac{d}{d x^0} = \frac{1}{N c} \frac{d}{dt} = \frac{1}{\tc} \frac{d}{d t} \,. \label{ddx0}
\end{align}
Here, it is important to note that \( \tc \) explicitly accounts for the time dependence of \( c \), incorporating its derivative. Moreover, in the meVSL model, \( \tc \) is assumed to take the form \cite{Lee:2020zts}
\begin{align}
\tc = c_0 a^{b/4} \quad , \textrm{where} \quad c_0 = \tc(a_0) \,\, \textrm{and} \,\, b = \textrm{const} \label{tc} \,.
\end{align}
$a_0$ is the current scale factor, and its value is $1$. Based on these, we obtain background Christoffel symbols as
\begin{equation}
	\bar\Gamma^0_{00} = 0 \quad , \quad \bar\Gamma^0_{ij} = \frac{1}{\tc} \dot{a} a \gamma_{ij} =  \frac{1}{\tc} a^2 H \gamma_{ij} \quad , \quad \bar\Gamma^i_{0j} = \frac{1}{\tc} \frac{\dot{a}}{a} \delta^i{}_j = \frac{1}{\tc} H \delta^i{}_j \;, \label{barGamma}
\end{equation}
where the dot denotes differentiation with respect to cosmic time, $t$.  The perturbed Christoffel symbols are given by
\begin{align}
	\delta \Gamma^0_{00} &= -\frac{1}{2 \tc} \dot{h}_{00} \;, \quad \delta \Gamma^0_{i0} = -\frac{1}{2} h_{00,i} + \frac{a}{\tc} H h_{0i} \;, \quad
	\delta \Gamma^i_{00} = \frac{\gamma^{ij}}{a^2} \left[ - \frac{1}{2} h_{00,j} + \frac{a}{\tc} \left( \dot{h}_{j0} + H h_{j0} \right) \right]  \;, \nonumber \\
	\delta \Gamma^0_{ij} &= -\frac{a}{2} \left( h_{0i,j} + h_{0j,i} \right) + \frac{a^2}{\tc} \left( \frac{1}{2} \dot{h}_{ij} + H h_{ij} + H h_{00} \gamma_{ij} \right) \;, \label{deltaGamma} \\
	\delta \Gamma^i_{j0} &=  \frac{\gamma^{ik}}{2a} \left( \frac{a}{\tc} \dot{h}_{jk} + h_{k0,j} - h_{j0,k} \right) \;, \quad
	\delta \Gamma^i_{jk} = \frac{\gamma^{il}}{2} \left( h_{lj,k} + h_{lk,j} - h_{jk,l} - 2 \frac{a H}{\tc} h_{l0} \gamma_{jk} \right)  \;. \nonumber
\end{align}

\subsubsection{The perturbed Ricci tensor and Einstein tensor}
\label{subsubsec:Ricci tensor} 

We compute the Ricci tensors using the previously derived Christoffel symbols \eqref{barGamma} and \eqref{deltaGamma}. Here, we ignore second and higher-order perturbation terms and consider only first-order perturbations. 
\begin{align}
	R_{\mu\nu} &\equiv \bar{R}_{\mu\nu} + \delta R_{\mu\nu} = \bar\Gamma^\rho_{\mu\nu,\rho} - \bar\Gamma^\rho_{\mu\rho,\nu}  + \bar\Gamma^\rho_{\mu\nu}\bar\Gamma^\sigma_{\rho\sigma} - \bar\Gamma^\rho_{\mu\sigma}\bar\Gamma^\sigma_{\nu\rho}\nonumber \\ 
	&+ \delta\Gamma^\rho_{\mu\nu,\rho} - \delta\Gamma^\rho_{\mu\rho,\nu} + \bar\Gamma^\rho_{\mu\nu}\delta\Gamma^\sigma_{\rho\sigma} + \delta\Gamma^\rho_{\mu\nu}\bar\Gamma^\sigma_{\rho\sigma} - \bar\Gamma^\rho_{\mu\sigma}\delta\Gamma^\sigma_{\nu\rho} - \delta\Gamma^\rho_{\mu\sigma}\bar\Gamma^\sigma_{\nu\rho}\;,
\end{align}\index{Ricci tensor!Perturbation}
Using the FLRW metric, the components of the background and first-order perturbed Ricci tensors are given as
\begin{align}
	\bar{R}_{00} &= \frac{3}{\tc^2} \left( - \dot{H} - H^2 + H \frac{\dot{\tc}}{\tc} \right) = \frac{3}{\tc^2} \left( - \frac{\ddot{a}}{a} + H \frac{\dot{\tc}}{\tc} \right)  = \bar{R}_{00}^{(\GR)} + 3 \frac{H^2}{\tc^2} \frac{d \ln \tc}{d \ln a} \nonumber \\ 
&\equiv \bar{R}_{00}^{(\GR)} + 3 \frac{H^2}{\tc^2} \frac{\tc'}{\tc} \;, \label{barR00App} \\ 
	\bar{R}_{ij} &= \frac{1}{\tc^2} g_{ij} \left(\frac{\ddot{a}}{a} + 2 H^2 - H \frac{\dot{\tc}}{\tc} \right) = \bar{R}_{ij}^{(\GR)} - \frac{g_{ij}}{\tc^2} H^2 \frac{\tc'}{\tc} \label{barRijApp} \;, \\
       \delta R_{00} &= -\frac{1}{2 a^2}\nabla^2h_{00} +\frac{1}{a \tc} \left( \dot{h}_{0k,}^{\,\,\,\,\,\,\,k} + H h_{0k,}^{\,\,\,\,\,\,k} \right) - \frac{3}{2} \frac{H}{\tc^2} \dot{h}_{00} - \frac{1}{2 \tc^2} \left( \ddot{h} + 2H \dot{h} \right) + \frac{H}{2 \tc^2} \dot{h} \frac{d \ln \tc}{d \ln a} \; \nonumber \\
				&\equiv \delta R_{00}^{(\GR)} + \frac{H^2}{2 \tc^2} h' \frac{\tc'}{\tc} \;, \label{deltaR00App} \\
	\delta R_{0i} &= \frac{a}{\tc^2} h_{0i} \left[ \frac{\ddot{a}}{a} + 2 H^2 - H^2 \frac{d \ln \tc}{d \ln a} \right] - \frac{H}{\tc} h_{00,i} +    \frac{1}{2\tc} \left( \dot{h}_{ik,}^{\,\,\,\,\,\,k} - \dot{h}_{,i} \right) + \frac{1}{2a} \left( h_{0k,i}^{\,\,\,\,\,\,\,\,\,k} - \nabla^2 h_{0i} \right) \; \nonumber \\
				&\equiv \delta R_{0i}^{(\GR)} - \frac{a}{\tc^2} H^2 \frac{\tc'}{\tc} h_{0i} \;, \label{deltaRi0App} \\
	\delta R_{ij} &= \frac{1}{2}h_{00,ij} + \frac{a^2}{\tc^2} \left[ \frac{H}{2} \dot{h}_{00} + \left( \frac{\ddot{a}}{a} + 2 H^2 \right)h_{00} \right] \gamma_{ij} + \frac{1}{2} \left[ h_{kj,i}^{\,\,\,\,\,\,\,\,k} + h_{ik,j}^{\,\,\,\,\,\,\,\,k} - \nabla^2 h_{ij} - h_{,ij} \right]  \nonumber\\ 
				&+ \frac{a^2}{\tc^2} \left[ \frac{1}{2} \ddot{h}_{ij} + \frac{3}{2} H \dot{h}_{ij} + \left( \frac{\ddot{a}}{a} + 2H^2 \right) h_{ij} \right] + \frac{aH}{2 \tc} \left( \frac{a}{\tc} \dot{h} - 2 h_{0k,}^{\,\,\,\,\,\,\,\,k} \right)  \gamma_{ij} - \frac{a}{2 \tc} \left( \dot{h}_{0i,j} + \dot{h}_{0j,i} \right) \nonumber\\ 
				&- \frac{a H}{\tc} \left( h_{0i,j} + h_{0j,i} \right) -\frac{a^2}{\tc^2} H  \frac{d \ln \tc}{d \ln a} \left( \frac{1}{2} \dot{h}_{ij} + H h_{ij} + H h_{00} \gamma_{ij} \right) \nonumber \\ 
				&\equiv \delta R_{ij}^{(\GR)}-\frac{a^2 H^2}{\tc^2} \frac{\tc'}{\tc} \left( \frac{1}{2} h'_{ij} + h_{ij} + h_{00} \gamma_{ij} \right) \;, \label{deltaRijApp} 
\end{align}
where we use $\dot{H} = \ddot{a}/a - H^2$ and $h \equiv h_{\,\,k}^{k}$ denotes the trace of $h_{ij}$.  The natural logarithm of the scale factor, \(N \equiv \ln a \), is defined as the number of e-foldings. The derivative with respect to \( \ln a \) is denoted by a prime. Additionally, the Ricci tensors labeled with a superscript \( ^{(\GR)} \) represent the quantities obtained in the standard model, where the speed of light is constant. We also use $dN / dt = H$.

The background and perturbed Ricci scalars are obtained by contracting the previously derived Ricci tensors as 
\begin{align}
	R &= \bar{R} + \delta R = \bar{g}^{\mu\nu} \bar{R}_{\mu\nu} + \bar{g}^{\mu\nu}\delta R_{\mu\nu} + \delta g^{\mu\nu}\bar{R}_{\mu\nu} \;, \label{RApp} \\
	\delta R &= -\delta R_{00} + \frac{1}{a^2}\gamma^{ij}\delta R_{ij} - h_{00} \bar{R}_{00} - \frac{1}{a^2} \gamma^{ik} h_{kl} \gamma^{jl} \bar{R}_{ij}\;. \label{deltaR} 
\end{align}
From this, the Ricci scalar in the FLRW metric can be explicitly expressed as 
\begin{align}
	\bar{R} &= \frac{6}{\tc^2} \left( \frac{\ddot{a}}{a} + H^2 - H \frac{\dot{\tc}}{\tc} \right) \equiv \bar{R}^{(\GR)} - \frac{6}{\tc^2} H^2 \frac{d \ln \tc}{d \ln a} \;, \label{barRcomApp} \\
	\delta R &= \frac{1}{a^2} \nabla^2h_{00} + \frac{3}{\tc^2} \left[ H \dot{h}_{00} + 2 \left( \frac{\ddot{a}}{a} + H^2 \right) h_{00} \right] -\frac{2}{a \tc} \left( \dot{h}_{0k,}^{\,\,\,\,\,\, k} + 3 H h_{0k,}^{\,\,\,\,\,\, k} \right) \nonumber \\ 
	&+ \frac{1}{\tc^2} \left( \ddot{h} + 4 H \dot{h} \right) + \frac{1}{a^2} \left( h_{k\,\,,l}^{\,\,\,l\,\,\,k} - \nabla^2 h  \right) - \frac{H}{\tc^2} \frac{d \ln c}{d \ln a} \left( \dot{h} + 6  H h_{0}^{0} \right) \; \nonumber \\
	&\equiv \delta R^{(\GR)} - \frac{H}{\tc^2} \frac{d \ln c}{d \ln a} \left( \dot{h} + 6  H h_{0}^{0} \right)  \label{deltaRcomApp} \;.
\end{align}
For practical calculations, it is often convenient to express the Einstein tensor with mixed indices. The full decomposition of the Einstein tensor is given by
\begin{align}
	G^{\mu}{}_{\nu} &= g^{\mu\rho}R_{\rho\nu} - \frac{1}{2}\delta^{\mu}{}_{\nu}R = \bar{g}^{\mu\rho}\bar{R}_{\rho\nu} - \frac{1}{2}\delta^{\mu}{}_{\nu}\bar{R} + \bar{g}^{\mu\rho}\delta R_{\rho\nu} + \delta g^{\mu\rho}\bar{R}_{\rho\nu} - \frac{1}{2}\delta^{\mu}{}_{\nu}\delta{R} \nonumber \\
	&\equiv \bar{G}^{\mu}{}_{\nu} + \delta G^{\mu}{}_{\nu} = \kappa \left( \bar{T}^{\mu}_{\nu}  + \delta T^{\mu}_{\nu} \right)  \label{EFEs2} \;,
\end{align}
where $\bar{G}^{\mu}{}_{\nu}$ represents the background Einstein tensor, computed solely from the background metric \( \bar{g}_{\mu\nu} \). The Einstein gravitational constant $\kappa$ is defined as 
\begin{align} \kappa = \frac{8 \pi \tG}{\tc^4} \label{kappa} \,, \end{align}
which remains a constant in the meVSL model, even though both $\tG$ and $\tc$ evolve over cosmic time. Since $\kappa$ is constant in the meVSL model, there is no need to include variations in $\kappa$ (\textit{i.e.}, $\delta \kappa = 0$) in the above perturbed equations \cite{Lee:2020zts}.  In addition, $\delta G^{\mu}{}_{\nu}$ denotes the linearly perturbed Einstein tensor, which depends on both the background metric $\bar{g}_{\mu\nu}$ and the perturbation components $h_{\mu\nu}$ as
\begin{align}
\bar{G}^0{}_0 &= - \frac{3}{\tc^2} H^2 \equiv \bar{G}^{0(\GR)}_0 \label{barG00} \\
\bar{G}^i{}_i &= -\frac{1}{\tc^2} \left[ 2 \frac{\ddot{a}}{a} + H^2 - 2 H^2 \frac{d \ln \tc}{d \ln a} \right] \equiv \bar{G}^{i(\GR)}_i + \frac{2}{\tc^2} H^2 \frac{d \ln \tc}{d \ln a} \label{barGii} \\
\delta G^0{}_0 &= -3 \frac{H^2}{\tc^2} h_{00} + 2 \frac{H}{a\tc} h_{0k,}^{\,\,\,\,\,\,\,k} - \frac{H}{\tc^2} \dot{h} + \frac{1}{2a^2} \left( \nabla^2 h - h_{k\,\,,l}^{\,\,\,l\,\,\,k} \right) \equiv \delta G^{0(\GR)}_0  \label{deltaG00gen} \;, \\
\delta G^0{}_i &=  \frac{H}{\tc} h_{00,i} + \frac{1}{2a} \left( \nabla^2 h_{0i} - h_{0k,i}^{\,\,\,\,\,\,\,\,\,\,k} \right) + \frac{1}{2 \tc} \left( \dot{h}_{,i} - \dot{h}_{ik,}^{\,\,\,\,\,\,\,\,k} \right) \equiv \delta G^{0(\GR)}_i \label{deltaG0igen} \;, \\
\delta G^i{}_j &= \frac{1}{2a^2} \left( h_{00,j}^{\,\,\,\,\,\,\,\,\, i} - \delta^{i}_{j} \nabla^2 h_{00} \right) + \frac{1}{2 a^2} \left[ h_{kj,}^{\,\,\,\,\,\,\,\,ik} + h_{k,j}^{i\,\,\,\,\,\,\,k} -\nabla^2 h^{i}_{j} - h_{\,,j}^{\,\,i} + \left(  \nabla^2 h - h_{k\,\,,l}^{\,\,l\,\,\,\,\,k}  \right) \delta^{i}_{j} \right] \nonumber\\ 
	&+  \frac{1}{2\tc^2} \left[ \ddot{h}^{i}_{j} + 3 H \dot{h}^{i}_{j} - \left( \ddot{h} + 3 H \dot{h} + 2 H \dot{h}_{00} + 2 \left( 2 \frac{\ddot{a}}{a} + H^2 \right) h_{00}  \right) \delta^{i}_{j} \right]  \nonumber\\ 
	&- \frac{1}{2 \tc a} \left[ \dot{h}_{0\,\,\,,j}^{\,\,i} + \dot{h}_{0j,}^{\,\,\,\,\,\,\,i} + 2H\left( h_{0\,,j}^{\,i} + h_{0j,}^{\,\,\,\,\,\,i} \right) - 2 \left( \dot{h}_{0k,}^{\,\,\,\,\,\,\,k} + 2H h_{0k,}^{\,\,\,\,\,\,\,k}   \right) \delta_{j}^{i} \right] \nonumber \\
	&-\frac{H}{2 c^2} \frac{d \ln \tc}{d \ln a} \left[ \dot{h}_{j}^{i} - \left( \dot{h} + 4 H h_{00} \right) \delta^{i}_{j} \right] \nonumber \\
	&\equiv \delta G^{i(\GR)}_j  -\frac{H}{2 c^2} \frac{d \ln \tc}{d \ln a} \left[ \dot{h}_{j}^{i} - \left( \dot{h} + 4 H h_{00} \right) \delta^{i}_{j} \right] \;. \label{deltaGijgen}
\end{align}
In cosmological perturbation theory, the e-folding number, \( N \), provides a convenient and intuitive measure of the universe’s expansion. Since the number of e-foldings directly correlates with the scale factor’s evolution, it is an essential parameter for comparing theoretical predictions with observational data. By rewriting the governing equations for the Fourier modes of perturbations in terms of \( N \), one can more easily track the evolution of these modes and their relation to the horizon scale, given by 
\begin{align}
    c k = aH \quad , \quad l_{\textrm{h}} \equiv \frac{ck}{aH} \label{ckaH} \,.
\end{align}
The horizon scale is typically the scale at which causal interactions can occur, meaning it's the distance that light can travel during the expansion of the universe over a certain period. In this case, $c/H$ refers to the Hubble radius or the scale beyond which light or any signal cannot travel. When you multiply this by $k/a$, the result describes the scale of perturbations relative to the expansion of the universe. Therefore, $c k/(a H)$ is a quantity describing the relationship of physical scale and Hubble horizon in cosmological perturbations. Thus, this formulation allows for a clearer understanding of how different perturbation variables evolve with respect to the horizon, facilitating a more systematic and insightful analysis of cosmological perturbations.

\subsection{Perturbations of the Energy-momentum tensor}
\label{subsec:perturbations Tmunu}

The background energy-momentum tensor for a perfect fluid is given by
 \begin{align} \bar{T}_{\mu\nu} = \left( \bar{\rho} + \frac{\bar{P}}{\tc^2} \right) \bar{U}_{\mu} \bar{U}_{\nu} + \bar{P} \bar{g}_{\mu\nu} \,,  \label{barT} \end{align}
 where $\bar{\rho}$ represents the mass density and $\bar{P}$ is the isotropic pressure. The four-velocity of the background fluid, denoted as $U^{\mu}$, is given by 
\begin{align}
\bar{U}^{\mu} \equiv \frac{d x^{\mu}}{dt} = \left( \tc\,, 0 \right) \,, \label{barUmu}
\end{align}
where we use the fact that, in the FLRW metric, the proper time of comoving observers is identical to the cosmic time.  In the context of the meVSL framework, we consistently interpret $\bar{\rho}$ as the rest mass density and define the total energy density of the background fluid as $\bar{\rho} \tilde{c}^2$, where $\tilde{c}$ is the time-dependent speed of light that is the $0$-th component of the four-velocity as shown in Eq.~\eqref{barUmu}. Accordingly, the pressure term is expressed as $\bar{P}/\tilde{c}^2$ in the energy-momentum tensor to ensure dimensional consistency and to preserve the covariant conservation law $\nabla^\mu T_{\mu\nu} = 0$ even under the time variation of $\tilde{c}$. This formulation guarantees that both the background dynamics and the perturbed quantities remain internally consistent. In particular, the choice of $\bar{U}^\mu = ( \tilde{c}, 0, 0, 0 )$ as the four-velocity of comoving observers reflects the fact that proper time coincides with cosmic time in the FLRW metric. We emphasize that this prescription is crucial for maintaining the diagonal form of the background energy-momentum tensor and for correctly capturing the effects of the varying speed of light in both the Einstein equations and the linear perturbation theory.

In the unperturbed FLRW background, this energy-momentum tensor is diagonal, reflecting the homogeneity and isotropy of the universe at large scales.  However, when metric perturbations are introduced, the energy-momentum tensor also acquires additional contributions that account for deviations from the perfect fluid approximation. These perturbations introduce new components that describe various physical effects. Heat fluxes, $q^{\mu}$, represent the transport of energy due to non-equilibrium processes. In a perfect fluid, the energy flow is purely determined by the four-velocity of the fluid. However, in the presence of perturbations, an additional energy flux component appears, which leads to energy exchange between different regions of the fluid. Anisotropic stresses, $\pi_{\mu\nu}$, describe deviations from an isotropic pressure. While a perfect fluid is characterized by an isotropic pressure, perturbations introduce directional pressure differences. These anisotropic stresses influence the evolution of metric fluctuations and play a crucial role in determining the growth of cosmic structures, particularly in the presence of shear and tensor perturbations.  This effect becomes relevant in scenarios where deviations from a simple barotropic equation of state occur.  Bulk viscosity accounts for non-adiabatic pressure variations that occur in a fluid undergoing expansion or compression. In an expanding universe, the presence of bulk viscosity ($\pi = \pi^{\mu}_{\,\,\mu}$ \textit{i.e.}, trace of the anisotropic stresses) modifies the effective pressure, thereby affecting the dynamics of cosmological perturbations. Thus, in the presence of perturbations, the total energy-momentum tensor acquires additional terms and can be expressed in the perturbed form as
\begin{align}
T_{\mu\nu} &= \bar{T}_{\mu\nu} + \delta T_{\mu\nu} \nonumber \\
&= \bar{T}_{\mu\nu} + \delta \rho \bar{U}_{\mu} \bar{U}_{\nu} + \bar{\rho} \delta U_{\mu} \bar{U}_{\nu} +  \bar{\rho} \bar{U}_{\mu} \delta U_{\nu} + q_{\mu} \bar{U}_{\nu} + q_{\nu} \bar{U}_{\mu} + \bar{\theta}_{\mu\nu} \left( \delta P + \pi \right) + \bar{P} \delta \theta_{\mu\nu} + \pi_{\mu\nu} \nonumber \\
&, \textrm{where} \quad \bar{\theta}_{\mu\nu} \equiv \frac{\bar{U}_{\mu} \bar{U}_{\nu}}{\bar{\tc}^2} + \bar{g}_{\mu\nu} \quad , \quad \delta \theta_{\mu\nu} \equiv \frac{\bar{U}_{\mu} \delta U_{\nu}}{\bar{\tc}^2} + \frac{\delta U_{\mu} \bar{U}_{\nu}}{\bar{\tc}^2} - 2 \frac{\delta c}{\bar{c}} \bar{U}_{\mu} \bar{U}_{\nu} + \delta g_{\mu\nu} \,, \nonumber \\ 
&\textrm{and} \quad q_0 = \pi_{\mu 0} = 0 \label{Tmunu} \,.
\end{align}
To obtain the total energy-momentum tensor, we must first determine the perturbation of the four-velocity. This can be derived from the normalization condition of the four-velocity, which applies to the full metric, including perturbations. Additionally, we first define the three-vector components as follows
\begin{align}
\delta U_i \equiv a v_i \label{deltaUi} \,.
\end{align}
We obtain the following relation from the normalization condition of the four-velocity, considering terms up to first-order perturbations
\begin{align}
g_{\mu\nu} U^{\mu} U^{\nu} &= \bar{g}_{\mu\nu} \bar{U}^{\mu} \bar{U}^{\nu} = -\tc^2 \quad \Rightarrow \quad 2 \bar{g}_{\mu\nu} \bar{U}^{\mu} \delta U^{\nu} = - \delta g_{\mu\nu} \bar{U}^{\mu} \bar{U}^{\nu} \label{deltaUmu} \,.
\end{align}
From equations \eqref{barUmu} and \eqref{deltaUmu}, we can obtain $\delta U^0$ 
\begin{align}
\delta U^{0} = \frac{\tc}{2} \delta g_{00} = \frac{\tc}{2} h_{00} \label{deltaU0u} \,.
\end{align}
The covariant components of the perturbed four-velocity, $\delta U_{\mu}$, are obtained by lowering the indices of the total four-velocity using the total metric
\begin{align}
U_{\mu} = g_{\mu\nu} U^{\nu} \quad \Rightarrow \quad \delta U_{\mu} = \bar{g}_{\mu\nu} \delta U^{\nu} + \delta g_{\mu\nu} \bar{U}^{\nu} \label{deltaUmul} \,.
\end{align}
The contravariant components of the perturbed four-velocity can be obtained by raising the indices of $\delta U_{\nu}$ in Eq.~\eqref{deltaUmul} using the inverse background metric $\bar{g}^{\mu\nu}$
\begin{align}
\delta U^{\mu} &= \bar{g}^{\mu\nu} \delta U_{\nu} - \bar{g}^{\mu\nu} \delta g_{\nu\rho} \bar{U}^{\rho} \label{deltaUmuu} \,.
\end{align}
By using Eqs.~\eqref{deltaUi}, \eqref{deltaU0u}, \eqref{deltaUmul}, and \eqref{deltaUmuu}, we also obtain
\begin{align}
\delta U_{0} = \frac{\tc}{2} h_{00} \quad , \quad \delta U^{i} = \frac{\gamma^{ij}}{a} \left( v_{j} - \tc h_{0j} \right) \,. \label{deltaU0l}
\end{align}
From these,  we obtain the explicit forms of therms in Eq.~\eqref{Tmunu}
\begin{align}
&\bar{\theta}_{00} = 0 \quad ,  \quad \bar{\theta}_{ij} = \bar{g}_{ij} \quad , \quad \bar{\theta}_{0i} = 0 \label{bartheta} \\
&\delta \theta_{00} = -2 \frac{\delta c}{\tc} \quad , \quad \delta \theta_{0i} = a \left( h_{0i} - \frac{v_i}{\tc} \right) \quad , \quad \delta \theta_{ij} = \delta g_{ij} = a^2 h_{ij} \label{deltatheta} \,.
\end{align}
The individual components of the energy-momentum tensor can be explicitly expressed as 
\begin{align}
T_{00} &= \bar{\rho} \tc^2 + \delta \rho \tc^2 - \bar{\rho} \tc^2 h_{00} - 2 \bar{P} \frac{\delta c}{\bar{\tc}} \equiv T^{(\GR)}_{00} + 2  \bar{P} \frac{\delta c}{\bar{\tc}} \label{T00} \\
T_{0i} &= - \left( \bar{\rho} + \frac{\bar{P}}{\tc^2} \right) c \delta U_i - \tc q_i + \bar{P} a h_{0i} \equiv T^{0(\GR)}_i \label{T0i} \\
T_{ij} &= \bar{P} \bar{g}_{ij} + \left( \delta P + \pi \right) \bar{g}_{ij} + \bar{P} \delta g_{ij} + \pi_{ij} \label{Tij}  \equiv T^{(\GR)}_{ij} \,.
\end{align}
These components describe the distribution of energy, momentum, and stress within the system. In perturbation theory, it might be possible to introduce a perturbation \( \delta c(x^{\mu}) \) to the background speed of light \( \tc(t) \) when there exists an additional physical variation beyond the homogeneous background quantity. Specifically, we express the speed of light as  
\begin{align}
c(x^{\mu}) = \tc (t) + \delta c (x^{\mu}) \,, \label{cxmu}
\end{align}
where $\delta c(x^{\mu})$ represents a perturbation that depends on both space and time. We should consider this perturbation if it is not simply a background evolution due to cosmic expansion but rather an additional dynamical fluctuation. In other words, $\delta c$ should not be interpreted as a trivial shift in the background speed of light caused by a small perturbation in time $\delta t$ if it is a real physical degree of freedom. We will discuss this aspect of the meVSL model in more detail later.Additionally, by contracting the energy-momentum tensor with the inverse metric, we obtain the energy-momentum tensor with mixed indices as
\begin{align}
T^0{}_0 &= - \bar{\rho} \tc^2 - \delta \rho \tc^2 + 2  \bar{P} \frac{\delta c}{\bar{\tc}} \equiv T^{0(\GR)}_0 + 2  \bar{P} \frac{\delta c}{\bar{\tc}} \label{Tm00} \\
T^0{}_i &= \left( \bar{\rho} + \frac{\bar{P}}{\tc^2} \right) c \delta U_i + \tc q_i  \equiv T^{0(\GR)}_i \label{Tm0i} \\
T^i{}_j &= \bar{P} \delta^i{}_j + \left( \delta P + \pi \right) \delta^i{}_j + \pi^i{}_j \label{Tmij}  \equiv T^{i(\GR)}_j \,.
\end{align}

\subsection{Coordinates and Gauge transformations}
\label{subsec:CGT}

A gauge is a mapping that defines a correspondence between points on the physical manifold and those on the background manifold, allowing for a meaningful comparison of tensors on these two manifolds. Since perturbations are described using fixed background coordinates, it is essential to take into account their transformation properties.

\subsubsection{Coordinate transformation}
\label{subsubsec:CT}

To analyze the effect of a gauge transformation (GT) on the functional dependence of perturbative quantities, we consider an infinitesimal coordinate transformation (CT) that relates a given gauge $G$ to another gauge $\hat{G}$, expressed as 
\begin{align}
&x^{\mu}(q) \mapsto \hat{x}^{\mu} (q) \equiv x^{\mu}(q)+ \xi^{\mu}(q) \label{CT01} \,, \\
&\hat{x}^{0}(q) = x^{0}(q) + \xi^{0}(q) \quad \textrm{and} \quad \hat{t}(q) = t(q) + \alpha(q) \label{x0} \\
&\hat{x}^{i}(q) = x^{i}(q) + \xi^{i}(q) \quad \textrm{where} \quad \xi^i \equiv \partial^i \beta + \epsilon^{(\tv) i} \quad \zeta_i \equiv \gamma_{ij} \xi^{j} \label{xi} 
\end{align}
where $x^{\mu}$ denotes the background coordinates and $\xi^{\mu}$ is an arbitrary vector field known as the gauge generator.  The function $\alpha$ specifies how the constant-time hypersurfaces are defined in the new coordinate system. Meanwhile, $\xi^i$ determines how the spatial coordinates are mapped on these hypersurfaces. The spatial shift can be decomposed into a scalar part, $\beta$, and a divergenceless vector part, $\epsilon^{(\tv) i}$.The gauge generator $\xi^{\mu}$ must satisfy $| \xi^{\mu} | \ll |x^{\mu}|$ to ensure that perturbations remain small. To examine how $x^0$ transforms in the meVSL model, we consider Eq. \eqref{x0}. First, we assume that the speed of light $c$ behaves as a scalar field
\begin{align}
	c(x^{\mu}) = \hat{c}(\hat{x}^{\mu}) \label{cCT} \,.
\end{align}
We use the fact that the transformation of a scalar field at each point $q$ must be the same because it is just relabeling the coordinates. In perturbation theory, we can define a perturbation of $\tc$ as Eq.~\eqref{cxmu} when there is an additional physical variation beyond the background quantity. In the meVSL model, the speed of light is solely a function of cosmic time and does not exhibit any additional dynamic perturbations. Therefore, $\delta c(x^{\mu})$ in the meVSL model is a dependent physical quantity that can be eliminated simply by a redefinition of time (i.e., a gauge choice). Thus, we can eliminate $\delta \hat{c}(x^\mu)$ as a mere gauge choice through a redefinition of the time coordinate $\hat{t}$.
\begin{align}
\hat{c} (\,\hat{t}\,) \simeq  c (t) + \frac{d c}{dt} \Bigl|_{t} \alpha (x^{\mu}) \equiv c (t) + \delta \hat{c} \quad \Rightarrow \quad
\delta \hat{c} = \frac{d c}{dt} \alpha \label{deltac2} \,. 
	\end{align} 
From this, $x^0$ and $\hat{x}^0$ can be expressed up to first-order perturbations as follows
\begin{align}
x^{0}\left( x^{\mu} \right) &= c \left( t \right) t \equiv \bar{x}^{0}(t)  \label{x0t} \,, \\
\hat{x}^{0}\left( \hat{x}^{\mu} \right) &= \left[ \hat{c} \left(t + \alpha \right) \right] \left[t + \alpha\left( x^{\mu} \right) \right] = \left[ c(t) + \frac{d c}{dt} \alpha \right] \left[t + \alpha\left( x^{\mu} \right) \right] \nonumber \\ 
	&\simeq c t + \left( 1 + \frac{t}{\tc} \frac{d c}{dt}\right) c \alpha \equiv c t + N c \alpha \equiv x^{0}\left( x^{\mu} \right) + \xi^{0}(x^{\mu}) \label{hatx0t} \,,
\end{align}
where $ \xi^{0} \equiv N c \alpha = \tc \alpha$.  

In the Friedmann–Lemaître–Robertson–Walker metric, a VSL $\tilde{c}(t)$ reflects a change in the clock rate across hypersurfaces, described by the lapse function. This variation is not a dynamical field evolution but a consequence of coordinate choice, as the cosmic time coincides with the proper time of comoving observers due to the Weyl postulate. From an action principle including $\tilde{c}(t)$, we derive that $\tilde{c}$ does not possess independent dynamics but rather imposes a constraint on the scale factor $a(t)$, indicating that it is not an additional degree of freedom \cite{Lee:2025pdu}. This insight reframes the VSL concept as a manifestation of gauge freedom in GR, wherein physical laws remain invariant under smooth coordinate transformations. Consequently, the time redefinition leading to Eq.~\eqref{hatx0t} is to be understood as a background coordinate transformation that does not constrain the choice of perturbative gauge conditions. In particular, the longitudinal (Newtonian) gauge remains valid and consistent for scalar perturbations, as the variation in $\tilde{c}$ does not alter the structure of gauge-invariant variables or introduce new dynamical modes. Therefore, the gauge conditions such as $B = 0$ and $E = 0$ imposed in the perturbation analysis are preserved and remain well-defined under this formulation.

In the meVSL model, the speed of light $c$ is assumed to be a function of cosmic time, meaning that on constant-time hypersurfaces, it remains uniform. Thus, $c$ is a function of $t$ without explicit dependence on spatial coordinates $x^i$. While a coordinate transformation can introduce an apparent spatial dependence of $c$, this is merely a consequence of the choice of coordinates rather than a physical effect \cite{Lee:2024zcu}. The key point is that the physical interpretation of $c$ should be coordinate-independent. In the meVSL framework, its variation is intrinsically tied to the evolution of cosmic time rather than spatial position. Thus, despite the possibility of coordinate-induced variations, the fundamental assumption of the meVSL model is that the speed of light is solely a function of time, $c = c(t)$, rather than a general function $c(x^\mu)$. Thus, we can write the line element including the scalar mode perturbations as
\begin{align}
ds^2 = -\tc^2(t) \left( 1 + 2 \psi \right) dt^2 + 2 \tc(t) a(t) B_i dt dx^i + a^{2}(t) \left[ \left( 1 + 2 \phi \right) \gamma_{ij} + h_{ij} \right] dx^i dx^j \label{scalarperds2} \,.
\end{align}

\subsubsection{Gauge transformation of the metric}
\label{subsubsec:GT}
Geometrically, under a GT, a fixed point on the background manifold is mapped to a corresponding point on the physical manifold. As a result, the coordinates of the mapped point differ from those of the original point and are determined by the transformation law.  The gauge generator $\xi^{\mu}$ can be regarded as a vector field on the physical manifold, governing the transformation properties of perturbative quantities. The variation of perturbative quantities under a GT is governed by the Lie derivative of the metric along $\xi^{\mu}$, which encapsulates how the functional form of these quantities changes in response to the gauge choice. 
Since the spacetime interval remains invariant under the CT given by Eq.~\eqref{CT01}, we can express this invariance as
\begin{align}
ds^2 = g_{\mu\nu} (x) dx^{\mu} dx^{\nu} = \hat{g}_{\alpha\beta}(\hat{x}) d\hat{x}^{\alpha} d\hat{x}^{\beta} = \frac{\partial \hat{x}^{\alpha}}{\partial x^{\mu}} \frac{\partial \hat{x}^{\beta}}{\partial x^{\nu}} \hat{g}_{\alpha\beta}(\hat{x}) dx^{\mu}  dx^{\nu}  \label{ds2} \,.
\end{align}
Thus, the transformation of the metric components under a CT is given by 
\begin{align}
g_{\mu\nu} (x) 
&\overset{{\cal O} (1)}{\simeq} \hat{g}_{\mu\nu}(x) + \partial_{\alpha} g_{\mu\nu}(x) \xi^{\alpha} + \partial_{\mu} \xi^{\alpha} g_{\alpha\nu}(x) + \partial_{\nu} \xi^{\alpha} g_{\alpha\mu}(x) \label{gmunuTran} \\
&= \hat{g}_{\mu\nu}(x) + \nabla_{\mu} \xi_{\nu} + \nabla_{\nu} \xi_{\mu}  \,. \nonumber
\end{align}
This equation explicitly shows how the metric components transform under a coordinate change, reflecting the gauge freedom. Using Eqs. \eqref{gmunuwhmunu}, \eqref{CT01}, and \eqref{gmunuTran}, we can express each metric component in terms of first-order perturbations as follows
\begin{align}
&\hat{h}_{00} = h_{00} + 2  \frac{1}{\tc} \dot{\xi}^{0} = h_{00} + 2 \left[ \frac{\dot{\tc}}{\tc} \alpha + \dot{\alpha} \right]  \label{hatg00} \,, \\
&\hat{h}_{0i} = h_{0i} - \frac{a}{\tc} \gamma_{ij} \dot{\xi}^{j} + \frac{1}{a} \partial_i \xi^0 = h_{0i} - \frac{a}{\tc} \dot{\zeta}_i + \frac{\tc}{a} \partial_i \alpha \label{hatg0i} \,, \\
&\hat{h}_{ij} = h_{ij} - \frac{2}{\tc} H \xi^0 \gamma_{ij} - \gamma_{kj} \partial_i \xi^k - \gamma_{ik} \partial_j \xi^k = h_{ij} - 2 H \alpha \gamma_{ij} - \partial_i \zeta_j - \partial_j \zeta_i \label{hatgij} \,.
\end{align}
In the case of GR, where $N = 1$, Eq.~\eqref{hatg00} reduces to the standard GT result in GR, ensuring consistency with conventional formulations.

\subsubsection{Gauge transformation of the energy-momentum tensor}
\label{subsubsec:GTTmunu}
Similarly, by applying the same approach used for the metric transformation in equation \eqref{gmunuTran}, we can derive the transformation rules for the components of the energy-momentum tensor
\begin{align}
\hat{T}_{\mu\nu} (x) \simeq T_{\mu\nu}(x) - \partial_{\alpha} T_{\mu\nu}(x) \xi^{\alpha} - \partial_{\mu} \xi^{\alpha} T_{\alpha\nu}(x) - \partial_{\nu} \xi^{\alpha} T_{\alpha\mu}(x) \label{TmunuTran} \,.
\end{align}
Using equations \eqref{Tmunu}, \eqref{hatg00}, \eqref{hatg0i}, and \eqref{hatgij},  we derive the GT relations for the first-order perturbations of $\delta \rho$, $v_i$, and $\delta P$ as follows
\begin{align}
&\delta \hat{\rho} = \delta \rho + 3 H \left( 1 + \omega \right) \rho \alpha \label{hatdeltarho} \,, \\
&\hat{v}_i =  v_{i} + \frac{\tc^2}{a} \partial_i \alpha \label{hatvi} \,, \\
&\delta \hat{P} = \delta P - \dot{\bar{P}} \alpha \label{hatdeltaP} \,.
\end{align}
According to the Stewart-Walker lemma, perturbations of quantities that are either zero or constant in the background are automatically gauge-invariant \cite{Stewart:1974uz}. This means that the perturbations for quantities such as heat flux ($\hat{q}_i = q_i$), bulk viscosity ($\hat{\pi} = \pi$), and anisotropic stress ($\hat{\pi}_{ij} = \pi_{ij}$) are unaffected by GT.

\subsection{The Scalar-Vector-Tensor Decomposition}
\label{subsec:SVT}

The scalar-vector-tensor (SVT) decomposition, introduced by Lifshitz \cite{Lifshitz46}, is a systematic method for separating different types of tensor perturbations. This method is applicable to both the metric and the energy-momentum tensor. For a perturbed metric, it is expressed in terms of scalar components ($A, C$), a vector component ($B_i$), and a tensor component ($E_{ij}$). Using Helmholtz’s theorem, the decomposition allows for the extraction of additional scalar and vector components from the perturbations. Similarly, the spatial component of the perturbed four-velocity, denoted by ($v_i$), can also be decomposed into scalar and vector parts.

\subsubsection{Vector Decomposition}
Any spatial vector \( B_i \) can be uniquely decomposed into an irrotational (longitudinal) component \( B_i^\parallel \) and a divergenceless (solenoidal) component \( B_i^\perp \), such that
\begin{equation}
B_i = B_i^\parallel + B_i^\perp \,,  \label{Bi}   
\end{equation}
where these components satisfy the following conditions
\begin{equation}
\epsilon_{ijk} \partial^j B_k^\parallel = 0, \quad \partial^k B_k^\perp = 0 \,. \label{Bicond} 
\end{equation}
By applying Stokes' theorem, the irrotational part \( B_i^\parallel \) can be written as the gradient of a scalar field \( B \), leading to
\begin{equation}
B_i = \partial_i B + B_i^{(\tv)}, \quad \text{with} \quad \partial_i B^{(\tv) i} = 0 \label{Bi2} \,,
\end{equation}
where $B$ represents the scalar part of $B_i$, while $B_i^{(\tv)}$ is the pure vector perturbation.  Similarly, the spatial gauge generator $\zeta_i$ in \eqref{xi} and the velocity perturbation $v_i$​ given in \eqref{deltaUi} can also be decomposed into its longitudinal and solenoidal components as 
\begin{align}
\zeta_i = \partial_i \beta + \epsilon^{(\tv)}_i \quad , \quad v_i = \partial_i v + v_i^{(\tv)} \label{vi2} \,. 
\end{align}

\subsubsection{Tensor Decomposition}
Similarly, any spatial rank-$2$ tensor \( E_{ij} \) can be decomposed into its longitudinal, solenoidal, and tensor components
\begin{equation}
E_{ij} = E^\parallel_{ij} + E^\perp_{ij} + E^{(\tT)}_{ij} \label{Eij} \,.
\end{equation}
The components satisfy the following conditions
\begin{equation}
\epsilon^{jlk} \partial_l \partial_k E^\parallel_{ij} = 0, \quad
\partial^i \partial^j E^\perp_{ij} = 0, \quad
\partial^j E^{(\tT)}_{ij} = 0 \label{Eijcond} \,.
\end{equation}
The longitudinal and orthogonal parts can be further decomposed as
\begin{equation}
E^\parallel_{ij} = \left(\partial_i \partial_j - \frac{1}{3} \delta_{ij} \nabla^2 \right) 2 E ,
\quad
E^\perp_{ij} = \left( \partial_i E^{(\tv)}_j + \partial_j E^{(\tv)}_i \right), \quad \text{with} \quad \partial^i E^{(\tv)}_i = 0 \label{Eij2} .
\end{equation}
In this decomposition, \( E \) is a scalar function, and \( E^{(\tv)}_i \) is a divergenceless vector.  Now starting from equation.~\eqref{gmunuwhmunu} and applying the SVT decomposition, each component of the metric perturbation can be expressed as follows
\begin{align}
h_{00} = - 2 A \quad , \quad h_{0i} = B_i \quad , \quad h_{ij} = 2D \gamma_{ij} + E_{ij} \label{hmunuSVT} \,.
\end{align}
In this decomposition, we identify four scalar functions $\left(A, B, D, E \right)$ and two transverse vector components $\left( B_i^{(\tv)}, E_i^{(\tv)}\right)$. The transverse vectors satisfy a transversality condition, meaning each transverse vector contains two independent functions, contributing a total of four degrees of freedom. Additionally, the remaining two degrees of freedom correspond to the tensor components $E_{ij}^{(\tT)}$. Using the SVT decomposition \eqref{hmunuSVT}, we can express the GT of the metric components \eqref{hatg00}-\eqref{hatgij} and those of the energy-momentum tensor \eqref{T00}-\eqref{Tij} in a more structured form as 
\begin{align}
&\hat{A} = A - \frac{\dot{\tc}}{\tc} \alpha - \dot{\alpha}  \quad , \quad 
\hat{B} = B + \frac{\tc}{a} \alpha - \frac{a}{\tc} \dot{\beta} \label{hatAB} \\
&\hat{D} = D - H \alpha - \frac{1}{3} \nabla^2 \beta \quad , \quad
\hat{E} = E - \beta \label{hatDE} \\
&\hat{B}_{i}^{(\tv)} = B_{i}^{(\tv)} - \frac{a}{\tc} \dot{\epsilon}_{i}^{(\tv)} \quad , \quad
\hat{E}_{i}^{(\tv)} = E_{i}^{(\tv)} - \epsilon_{i}^{(\tv)} \quad , \quad
\hat{E}_{ij}^{(\tT)} = E_{ij}^{(\tT)} \label{hatEij} \\
&\delta \hat{\rho} = \delta \rho + 3 H \left( 1 + \omega \right) \rho \alpha \quad , \quad \delta \hat{P} = \delta P - \dot{\bar{P}} \alpha \label{hatrhoP} \\
&\hat{v} = v + \frac{\tc^2}{a} \alpha \quad , \quad \hat{v}_{i}^{(\tv)} = v_i^{(\tv)} \label{hatv} \,. 
\end{align}

\subsection{Normal Mode Decomposition}

In this section, we explicitly formulate the Einstein equations by selecting an appropriate gauge with normal mode decomposition. Since perturbation equations are second-order partial differential equations, normal mode decomposition simplifies their treatment by transforming them into algebraic equations in Fourier space. This approach offers several advantages in cosmological perturbation theory. First, it transforms differential equations into algebraic equations, simplifying their solution. Second, it allows for a straightforward analysis of perturbations on different length scales. Third, it facilitates direct comparison with observational data, often analyzed in Fourier space. By adopting this formalism, we gain deeper insight into the evolution of perturbations and their role in the dynamics of the early universe.

To facilitate this decomposition, we introduce the eigenmodes \( Q(\mathbf{k}, \mathbf{x}) \) of the Laplacian operator, satisfying the Helmholtz equation
\begin{equation}
\nabla^2 Q(\mathbf{k}, \mathbf{x}) = -k^2 Q(\mathbf{k}, \mathbf{x}) \quad \,, \quad Q(\mathbf{k}, \mathbf{x}) = e^{i \mathbf{k} \cdot \mathbf{x}} \label{NMD} \,,
\end{equation}
where these eigenmodes correpond to plane waves for flat spatial slicing. Given a perturbation quantity \( h(t, \mathbf{x}) \), its Fourier mode $\check{h}(t, \mathbf{k})$ decomposition is given by the forward and inverse Fourier transforms
\begin{equation}
\check{h}(t, \mathbf{k}) = \int d^3x \ h(t, \mathbf{x}) e^{-i \mathbf{k} \cdot \mathbf{x}} \quad , \quad
h(t, \mathbf{x}) = \int \frac{d^3k}{(2\pi)^3} \check{h}(t, \mathbf{k}) e^{i \mathbf{k} \cdot \mathbf{x}} \label{FT} \,.
\end{equation}
This transformation converts differential equations in real space into algebraic equations in Fourier space, simplifying the analysis of perturbations at different scales. Therefore, using the SVT decomposition given by equations \eqref{Bi2}, \eqref{vi2}, and \eqref{Eij}, the equations for the Fourier modes of the perturbation variables can be rewritten as follows
\begin{align}
&\check{B}_i (t, \mathbf{k}) = i k_i \check{B} (t, \mathbf{k}) + \check{B}_i^{(\tv)} (t, \mathbf{k}) \quad ,  \quad 
\check{v}_i  (t, \mathbf{k})  =  i k_i \check{v}  (t, \mathbf{k}) + \check{v}_i^{(\tv)}  (t, \mathbf{k}) \,, \label{vi2FT} \\
&\check{E}^\parallel_{ij}(t, \mathbf{k}) = - \left(k_i k_j - \frac{1}{3} \delta_{ij} k^2 \right) 2 \check{E}(t, \mathbf{k}) \quad ,  \quad 
\check{E}^\perp_{ij}(t, \mathbf{k}) = i k_i \check{E}^{(\tv)}_j (t, \mathbf{k}) + i k_j \check{E}^{(\tv)}_i (t, \mathbf{k}) \,. \label{EperFT} 
\end{align}

\section{Einstein field equations}
\label{sec:EFEs}

The Einstein field equation (EFE) is expressed as
\begin{align}
G_{\mu\nu} + \Lambda g_{\mu\nu} = \frac{8 \pi \tG}{\tc^4} T_{\mu\nu} \equiv \kappa T_{\mu\nu} \label{EFEs} \,,
\end{align}
where $G_{\mu\nu}$ is the Einstein tensor, which encapsulattes the curvature of spacetime due to the presence of matter and energy. The cosmological constant $\Lambda$ accounts for the late time accelerated expansion of the Universe, while $T_{\mu\nu}$ represents the energy-momentum tensor describing the distribution of matter and radiation. When considering small deviations from the homogeneous and isotropic background, we introduce perturbations into the EFE. 

\subsection{For background}
\label{subsec:BG}

From Eqs.~\eqref{barG00}, \eqref{barGii}, \eqref{Tm00}, and \eqref{Tmij}, we derive the background Friedmann equations within the framework of the meVSL model, which governs the evolution of the Hubble parameter $H$, the acceleration of the scale factor $\ddot{a}/a$, and the energy densities $\bar{\rho}_i$​ of different cosmic components, including radiation, matter, and the cosmological constant $\Lambda$
\begin{align} &H^2 = \frac{8 \pi \tG}{3} \sum_{l} \bar{\rho}_l = \left[ \frac{8 \pi \tG_0}{3} \sum_{l} \bar{\rho}_{l0} a^{-3(1+\omega_l)} \right] a^{\frac{b}{2}} ,, \quad \textrm{where} ,, \bar{\rho}_{\Lambda} \equiv \frac{\Lambda \tc^2}{8 \pi \tG} \label{H} ,, \\ 
&\frac{\ddot{a}}{a} = - \frac{4 \pi \tG}{3} \sum_{l} \left( 1 + 3 \omega_l \right) \bar{\rho}_l + H^2 \frac{d \ln \tc}{d \ln a} \label{ddotaoa} ,. \end{align}
The meVSL model introduces a modification to the standard Friedmann equation by incorporating an additional factor $a^{b/2}$, which alters the evolution of the Hubble parameter. Moreover, the acceleration equation gains an extra term proportional to $H^2 d \ln \tc / d \ln a$, reflecting the influence of the varying speed of light on the cosmic expansion rate.
The evolution equation for the energy densities of cosmic components is given by
\begin{align} \frac{d \ln \bar{\rho}_i}{d \ln a} + 3 \left( 1 + \omega_i \right) + 2 \frac{d \ln \tc}{d \ln a} = 0 \quad \Rightarrow \quad \bar{\rho}_l = \bar{\rho}_{i0} a^{-3 \left( 1 + \omega_l \right) -\frac{b}{2}} ,, \label{barrhoi} \end{align}
where we assume that the speed of light evolves as $\tc = \tc_0 a^{b/4}$. This modification affects the dilution rates of different cosmic fluids, leading to deviations from standard cosmological models. As a result, this modification leads to a change in the rest mass while ensuring the validity of mass-energy equivalence within the meVSL model \cite{Lee:2020zts}.

\subsection{For scalar perturbations}
\label{subsec:SP}

The Einstein field equations for scalar perturbations play a crucial role in understanding the evolution of cosmic structures and the dynamics of gravitational potentials in an expanding universe. These equations describe how matter density fluctuations and metric perturbations interact, governing the formation of large-scale structures and the anisotropies observed in the cosmic microwave background. By relating the perturbations in the metric to those in the energy-momentum tensor, they provide a framework for analyzing the growth of perturbations and the effects of modified gravity or varying fundamental constants. In the context of the meVSL model, where the speed of light varies over cosmic time, these equations must be carefully re-examined to assess their implications for structure formation and cosmological observations.

We discuss scalar mode perturbations and derive the corresponding equations using the Newtonian gauge ($A \equiv \psi, D \equiv \phi, B = E = 0$). This gauge is particularly advantageous as it provides an intuitive interpretation of gravitational potentials ($\psi\,, \phi$), directly linking them to density perturbations and the motion of matter. Newtonian potential \(\Psi\) affects the time-time component \( g_{00} \) of the metric and is analogous to the gravitational potential arising from mass in Newtonian mechanics. Curvature potential \(\Phi\) influences the spatial components \( g_{ij} \) of the metric, representing changes in the spatial curvature structure. By eliminating non-physical gauge degrees of freedom, the Newtonian gauge simplifies the equations while maintaining a clear physical meaning, making it especially useful for analyzing cosmological structure formation. From equations \eqref{deltaG00gen}, \eqref{deltaG0igen}, \eqref{deltaGijgen}, \eqref{Tm00}, \eqref{Tm0i}, \eqref{Tmij},
\begin{align}
\delta G^{0}_{\,\,0} &: -6 \frac{H}{\tc^2} \dot{\phi} + 6 \frac{H^2}{\tc^2} \psi - 2 \frac{k^2}{a^2} \phi = - \frac{8 \pi \tG}{\tc^2}  \sum_{l} \delta \rho_{l} \nonumber \\
&: \phi' - \psi + \frac{1}{3} l_{\textrm{h}}^2 \phi = \frac{1}{2} \sum_l \Omega_l \delta_l \label{deltaG00} \,, \\
\delta G^{0}_{\,\,i} &: i k_i \frac{2}{\tc} \left( \dot{\phi} - H \psi  \right) = \frac{8 \pi \tG}{\tc^4} \sum_{l} \left( 1 + \omega_{l} \right)  \bar{\rho}_{l}  \tc a i k_i v_l  \nonumber \\
&: \phi' - \psi = \frac{3}{2} \frac{1}{l_{\textrm{h}}} \sum_{l} (1 + \omega_l ) \Omega_l \frac{\theta_l}{\tc} \label{deltaG0i} \,, \\
\delta G^{l}_{l} &: \frac{6}{\tc^2} \bigg[ - \ddot{\phi} + H \left( \dot{\psi} - 3 \dot{\phi} \right) + \left( 2 \frac{\ddot{a}}{a} + H^2 \right) \psi - \frac{ k^2 \tc^2}{3 a^2} \left(\psi + \phi \right) \nonumber \\ &+ \frac{\dot{\tc}}{\tc} \left( \dot{\phi} - 2H \psi \right) \bigg] = \frac{24 \pi \tG}{\tc^4} \sum_{l} \delta P_{l} \label{deltaGll}\\
&: \phi'' + \left( 2 + q \right) \phi' - \psi' - \left( 1 + 2 q \right) \psi + \frac{l_{\textrm{h}}^2}{3} \left( \psi + \phi \right) - \frac{\tc'}{\tc} \left( \phi' - 2 \psi \right) = \frac{3}{2} \sum_{l} \frac{c_{sl}^2}{\tc^2} \Omega_{l} \delta_l \nonumber   \,, \\
\delta G^{i}_{j} &: \frac{k^2}{a^2} \left( \phi + \psi \right) = \frac{12 \pi \tG}{\tc^4} \hat{k}_i \hat{k}_j \sum_l \pi^{ij}_{l} \quad \Rightarrow \quad l_{\textrm{h}}^2 \left( \phi + \psi \right) = \frac{9}{2} \sum_l (1 + \omega_l) \Omega_l \sigma_l \label{deltaGij} \,,
\end{align}
where $\Omega_l$ is the density parameter of the $l$-th component and $\delta_l$ is its density contrast, $\theta_l = k v_l$, and $q = - \ddot{a}/(a H^2)$ is the deceleration parameter as defined in
\begin{align}
q = -\frac{1}{2} \sum_{l} \left( 1 + 3 \omega_l \right) \Omega_{l} + \frac{\tc'}{\tc} \label{q} \,.
\end{align}
We also use the definition of shear scalar $\sigma$ as a specific combination of the anisotropic stress tensor, given by $
\sigma_l = - k^i k^j \pi_{ij}/(\rho_l \tc^2+ P_l)$ \cite{Ma:1995ey}. The anisotropic stress tensor $\pi_{ij}$ arises from the quadrupole moments of photons and neutrinos and possesses a traceless property.
It implies that only one gravitational potential is possible without a quadrupole moment present. In the early universe, neutrinos prevent this equality during CDM dominance. Similarly, if CDM or DE dominates but gravity deviates from GR, one might still find \(\phi \neq -\psi\).

The conservation of energy-momentum follows directly from the Einstein equations. Then, the perturbed energy-momentum conservation equations lead to the continuity and the Euler equation 
\begin{align}
&\dot{\delta}_{l}= - 3H \left( \frac{\tc_{ls}^2}{\tc^2} - \omega_l \right) \delta_l - \left( 1 + \omega_l \right) \left( \frac{\theta_l}{a} + 3 \dot{\phi} \right)  \label{dotdeltaNG2} \,, \\
&\dot{\theta}_l = H \left(3 \omega_l - 1 \right) \theta_l - \frac{\dot{\omega}_l}{1+\omega_l} \theta_l + \frac{\tc_{ls}^2 k^2}{(1+\omega_l) a} \delta_l + \frac{\tc^2 k^2}{a} \psi - \frac{\tc^2 k^2}{a} \sigma  + \frac{\dot{\tc}}{\tc} \theta_l \label{dotthetaNG2} \,.
\end{align}
We can rewrite the above equation as a function of the number of e-folding   
\begin{align}
&\delta_l'= - 3 \left( \frac{\tc_{ls}^2}{\tc^2} - \omega_l \right) \delta_l - \left( 1 + \omega_l \right) \left( \frac{\theta}{aH} + 3 \phi' \right)  \label{dotdeltaNGn} \,, \\
&\theta_l' = - \left( 1 - 3\omega_l \right) \theta_l - \frac{\omega_l'}{1 +\omega_l} \theta_l + \frac{1}{(1+\omega_l)}\frac{\tc_{ls}^2 k^2}{a H} \delta_l + \frac{\tc^2 k^2}{a H} \psi - \frac{\tc^2 k^2}{aH} \sigma  + \frac{\tc'}{\tc} \theta_l \label{dotthetaNGn} \,.
\end{align}

\begin{itemize}
	\item Cold dark matter (CDM): \\
CDM interacts with other particles solely through gravity, which is modeled as a pressureless perfect fluid. However, the CDM fluid velocity is generally non-zero in the conformal Newtonian gauge. In $k$-space, equations \eqref{dotdeltaNGn} and \eqref{dotthetaNGn} yield
\begin{align}
&\delta_c'= - \frac{\theta_c}{aH} - 3 \phi' \,,  \label{dotdeltaNGnCDM} \\
&\theta_c' = - \theta_c - \frac{\tc^2 k^2}{a H} \phi + \frac{b}{4} \theta_c \label{dotthetaNGnCDM} \,.
\end{align}
If we combine two equations \eqref{dotdeltaNGnCDM} and \eqref{dotthetaNGnCDM}, we obtain
\begin{align}
&\delta_c'' + \left( 2 + \frac{\dot{H}}{H^2} - \frac{\tc'}{\tc} \right) \delta_c' = -\frac{\tc^2k^2}{a^2 H^2} \psi - 3 \phi'' -3 \left(2 + \frac{\dot{H}}{H^2} - \frac{\tc'}{\tc}  \right)  \phi'  \nonumber \\
&, \textrm{where} \quad \frac{\dot{H}}{H^2} - \frac{\tc'}{\tc} = -\frac{3}{2} \sum_l \left( 1 + \omega_l \right) \Omega_l  \label{ddotdeltaNGnCDM}
\end{align}
	\item Baryon-photon fluid: \\
For the baryon and photon, we need to include the Thomson scattering term in their Euler equations~\eqref{dotthetaNG2} 
\begin{align}
&\dot{\theta}_{b} = -H \theta_{b} + \frac{\tc_{b}^2 k^2}{a} \delta_{b} + \frac{\tc^2 k^2}{a} \psi -\frac{2k^2}{5} \Theta_{2}^{b} + \frac{\dot{\tc}}{\tc} \theta_{b} + \frac{\dot{\tau}}{R} \left(\theta_b - \theta_{\gamma} \right) \label{dotthetabNG} \,, \\
&\dot{\theta}_{\gamma} =  \frac{3}{4} \frac{\tc_{\gamma}^2 k^2}{a} \delta_{\gamma} -\frac{2k^2}{5} \Theta_{2}^{\gamma} + \frac{\tc^2 k^2}{a} \psi + \frac{\dot{\tc}}{\tc} \theta + \dot{\tau} \left(\theta_b - \theta_{\gamma} \right)  \label{dotthetagammaNG}  \,,
\end{align}
where $\Theta_2$s are quadrupole, $R \equiv 3 \bar{\rho}_b/(4 \bar{\rho}_{\gamma})$, and the optical depth $\dot{\tau} \equiv n_e \sigma_{T}$ with $n_e$ being the number density of free electrons and $\sigma_T$ the Thomson cross-section. Before recombination, photons and baryons are tightly coupled, and we consider the combined photon-baryon system as a single imperfect fluid (\textit{i.e.}, tightly coupled approximation (TCA)). In this context, although the baryonic sound speed \( c_b \) is assumed to vanish (i.e., \( c_b = 0 \)) due to the pressureless nature of baryons, we do not directly evolve baryonic perturbations \( \delta_b \) and \( \theta_b \) independently. Instead, we utilize the effective plasma variables \( \delta_x \) and \( \theta_x \), which encode the collective behavior of the photon-baryon fluid \cite{Ma:1995ey,Pitrou:2010ai}. In the TCA,  we can put $\theta_{\gamma} \approx \theta_{b} \equiv \theta_x$ and use the adiabatic initial condition $3 \delta_{\gamma} = 4 \delta_{b}$.  Then the continuity equation of $\gamma$ is exactly equals to that of $b$ as
\begin{align}
\delta_{\gamma}' = -\frac{4}{3} \frac{\theta_{x}}{aH} - 4 \phi' \quad \Longleftrightarrow \quad
\delta_{b}' = - \frac{\theta_{x}}{aH} - 3 \phi' \label{deltagammab} \,.
\end{align} We also combine the Euler equations of $\gamma$ and $b$ to obtain the Euler equation for $\theta_{x}$ as
\begin{align}
\theta_{x}' = -\frac{R}{1+R} \theta_x + \frac{\tc_x^2 k^2}{aH} \left( \delta_b -  \phi \right) + \frac{b}{4} \theta_x \label{thetaxprime} \,.
\end{align}
We regard the plasma of photons and baryons as an imperfect fluid. Specifically, the background energy density and pressure of the plasma are given by
\begin{equation}
\bar{\rho}_{x} \equiv \bar{\rho}_{b} + \bar{\rho}_{\gamma} \quad , \quad \frac{\bar{P}_{x}}{\tc^2} \equiv \omega_x \bar{\rho}_{x} = \frac{1}{3} \bar{\rho}_{\gamma} \label{barrhoxPx} \,.
\end{equation}
We can also define the total mass density and pressure of the plasma as
\begin{align}
&\rho_{x} = \bar{\rho}_x + \delta \rho_x = \bar{\rho}_b + \delta \rho_b + \bar{\rho}_{\gamma} + \delta \rho_{\gamma} \label{rhox} \,, \\
&P_{x} = \bar{P}_x + \delta P_{x} = P_{\gamma} \label{Px} \,.
\end{align}
The equation of state and the adiabatic sound speed of the plasma can be inferred by analyzing its time evolution as
\begin{equation}
\omega_{x} = \omega_{\gamma} \frac{\bar{\rho}_{\gamma}}{\bar{\rho}_{x}} = \frac{1}{(3 + 4R)} \quad , \quad \tc_{x}^2 \equiv \frac{\dot{\bar{P}}_x}{\dot{\bar{\rho}}_x} = \frac{1}{3(1+R)} \left[ 1 + \frac{b}{4} \right] \tc^2 \, \label{omegaxcx2} \,,
\end{equation}
which incorporates the correction due to the varying speed of light (VSL) effect parameterized by \( b \). Thus, the sound speed \( \tc_x \) is the physically meaningful quantity that governs acoustic oscillations in the tight-coupling regime, and not \( \tc_b \) or \( \tc_\gamma \) independently. To avoid confusion, we clarify that \( \tc_b = 0 \) is a standard assumption in Boltzmann codes, while \( \tc_\gamma^2 = \frac{1}{3} \tilde{c}^2 \) and \( \tc_x^2 \) in Eq.~\eqref{omegaxcx2} reflect the total fluid behavior.  

	\item Curvature potential: \\
We can rewrite the Eq.~\eqref{deltaG00} with adopting two-fluid approximation
	\begin{align}
	\phi' \approx -\phi - \frac{l_{\textrm{h}}^2}{3}  \phi + \frac{1}{2} \left(  \Omega_{\gamma} \delta_{\gamma} + \Omega_b \delta_b + \Omega_{c} \delta_c \right) = -\phi - \frac{l_{\textrm{h}}^2}{3} \phi + \frac{1}{2} \left( \frac{1+R}{R} \Omega_{b} \delta_b + \Omega_{c} \delta_c \right)   \label{deltaG00TFA} \,.
	\end{align}
\end{itemize} 
We now numerically solve tightly coupled approximation equations \eqref{dotdeltaNGnCDM}, \eqref{dotthetaNGnCDM}, \eqref{deltagammab}, \eqref{thetaxprime}, and \eqref{deltaG00TFA} to investigate the impact of the speed of light variation on the dark matter density and the curvature potential. 
\begin{align}
&\delta_c'= - \frac{\theta_c}{aH} - 3 \phi' \quad , \quad \theta_c' = - \theta_c - \frac{\tc^2 k^2}{a H} \phi + \frac{b}{4} \theta_c \nonumber \\
&\delta_{b}' = -\frac{\theta_{x}}{aH} - 3 \phi' \quad , \quad 
\theta_{x}' = -\frac{R}{1+R} \theta_x + \frac{\tc_x^2 k^2}{aH} \left( \delta_b -  \phi \right) + \frac{b}{4} \theta_x \label{TCA} \\
&\phi' = - \left( 1 + \frac{l_{\textrm{h}}^2}{3} \right) \phi + \frac{1}{2} \left( \frac{1+R}{R} \Omega_{b} \delta_b + \Omega_{c} \delta_c \right) \nonumber
\end{align}
For the numerical evolution, we adopt adiabatic initial conditions for the perturbation variables, consistent with observations. At early times (specifically, in the deep radiation-dominated era and on super-horizon scales), the initial value for the gravitational potential is set to $\phi(n_i) = 10^{-5}$. Correspondingly, the initial conditions for the other fundamental variables in the Newtonian conformal gauge are set as $\delta_c(n_i) = -\frac{3}{2}\phi(n_i)$, $\theta_c(n_i) \approx 0$, $\delta_x(n_i) = -2\phi(n_i)$, and $\theta_x(n_i) \approx 0$. These initial values ensure consistency with the established adiabatic paradigm where velocity perturbations are negligible on super-horizon scales and the matter and radiation density perturbations are proportional to the gravitational potential. For the background cosmological model, we utilize the fiducial parameters from Planck 2018 (DR3) results \cite{Planck:2018vyg}, specifically: the physical baryon density $\Omega_b h^2 = 0.02237$, the physical cold dark matter density $\Omega_c h^2 = 0.1200$, the present-day Hubble constant $H_0 = 67.36 \text{ km s}^{-1} \text{ Mpc}^{-1}$, the dark energy density $\Omega_{\Lambda} \approx 0.6853$ (derived from $\Omega_m + \Omega_\Lambda = 1$ for a flat universe), the photon density $\Omega_{\gamma} h^2 \approx 2.47 \times 10^{-5}$, the effective number of relativistic species $N_{\text{eff}} = 3.046$, from which the total neutrino density $\Omega_{\nu} h^2 \approx N_{\text{eff}} \times (7/8) \times (4/11)^{4/3} \Omega_{\gamma} h^2 \approx 1.70 \times 10^{-5}$ is obtained (assuming massless neutrinos in the radiation component), leading to a total radiation density $\Omega_{\text{r}} h^2 \approx 4.17 \times 10^{-5}$. The scale factor at matter-radiation equality is given by $z_{\text{eq}} \approx 3414$ (\textit{i.e.}, $n_{\text{eq}} \approx -8.14$).

\subsection{The implications of the meVSL model}
\label{subsec:deltacphi}
We can summarize the physical Interpretation of each equation in equation.~\eqref{TCA}. 
\begin{itemize}
    \item \textbf{$\delta_c'$ (evolution of CDM density perturbations):}
    \begin{itemize}
        \item $-\frac{\theta_c}{aH}$: This term describes the contribution of CDM velocity perturbations ($\theta_c$) to density perturbations, influenced by the gravitational potential. 
        \item $-3\phi'$: This term accounts for the growth of density perturbations due to changes in the gravitational potential ($\phi$), incorporating the dilution effect caused by cosmic expansion.
        \item CDM, lacking pressure, is primarily governed by gravity. Consequently, the growth of its density perturbations is directly dependent on the gravitational potential and velocity perturbations.
    \end{itemize}

    \item \textbf{$\theta_c'$ (evolution of CDM velocity perturbations):}
    \begin{itemize}
        \item $-\theta_c$: This term represents the dilution of velocity due to cosmic expansion; velocities naturally decay in an expanding universe.
        \item $-\frac{\tc^2k^2}{aH}\phi$: This term signifies the change in velocity induced by the gradient of the gravitational potential, reflecting the gravitational pull on matter.
        \item $+\frac{b}{4}\theta_c$: This is a characteristic term of the meVSL model, describing the effect of varying light speed ($c = c_0 e^{(b/4)n}$) on velocity perturbations. If $b < 0$, implying a faster speed of light in the past, this term can either slow down the decay or accelerate the growth of velocity perturbations.
    \end{itemize}

    \item \textbf{$\delta_b'$ (evolution of baryon density perturbations):}
    \begin{itemize}
        \item $-\frac{\theta_x}{aH}$: This term indicates the contribution of baryon (and photon) velocity perturbations ($\theta_x$) to density perturbations, influenced by the gravitational potential. In the TCA regime, baryons move cohesively with photons.
        \item Similar to dark matter, baryon density perturbations evolve under gravity; however, due to interactions with photons, the velocity term is denoted as $\theta_x$.
    \end{itemize}

    \item \textbf{$\theta_x'$ (evolution of baryon-photon fluid velocity perturbations):}
    \begin{itemize}
        \item $-\frac{R}{1+R}\theta_x$: This term accounts for the dilution of velocity due to cosmic expansion and interactions with photons. $R = \frac{3\rho_b}{4\rho_\gamma}$ represents the ratio of baryon to photon energy densities.
        \item $+\frac{\tilde{c}_x^2 k^2}{aH} (\delta_b - \phi)$: This term describes the generation of velocity perturbations by the pressure gradient of sound waves, where $\tilde{c}_x$ is the sound speed. As the baryon-photon fluid possesses pressure, this term is crucial.
        \item $+\frac{b}{4}\theta_x$: This is a characteristic term of the meVSL model, representing the impact of varying light speed on baryon-photon fluid velocity perturbations.
    \end{itemize}

    \item \textbf{$\phi'$ (evolution of the Newtonian potential):}
    \begin{itemize}
        \item $- \left( 1 + \frac{l_{\textrm{h}}^2}{3} \right) \phi$: These terms represent the intrinsic evolution of the gravitational potential and contributions related to the horizon scale ($l_h$).
        \item $+\frac{1}{2}\left(\frac{1+R}{R}\Omega_b\delta_b + \Omega_c\delta_c\right)$: The gravitational potential is generated by the density perturbations of matter. This term illustrates how perturbed matter densities influence the gravitational potential.
        \item The gravitational potential is determined by the total energy density perturbations of the universe. Therefore, both dark matter and baryon density perturbations contribute to the evolution of $\phi$.
    \end{itemize}
\end{itemize}

\subsubsection{For a constant speed of light (\textit{i.e.},  $b=0$)}
\begin{figure*}
\centering
\vspace{1cm}
\begin{tabular}{cc}
\includegraphics[width=0.47\linewidth]{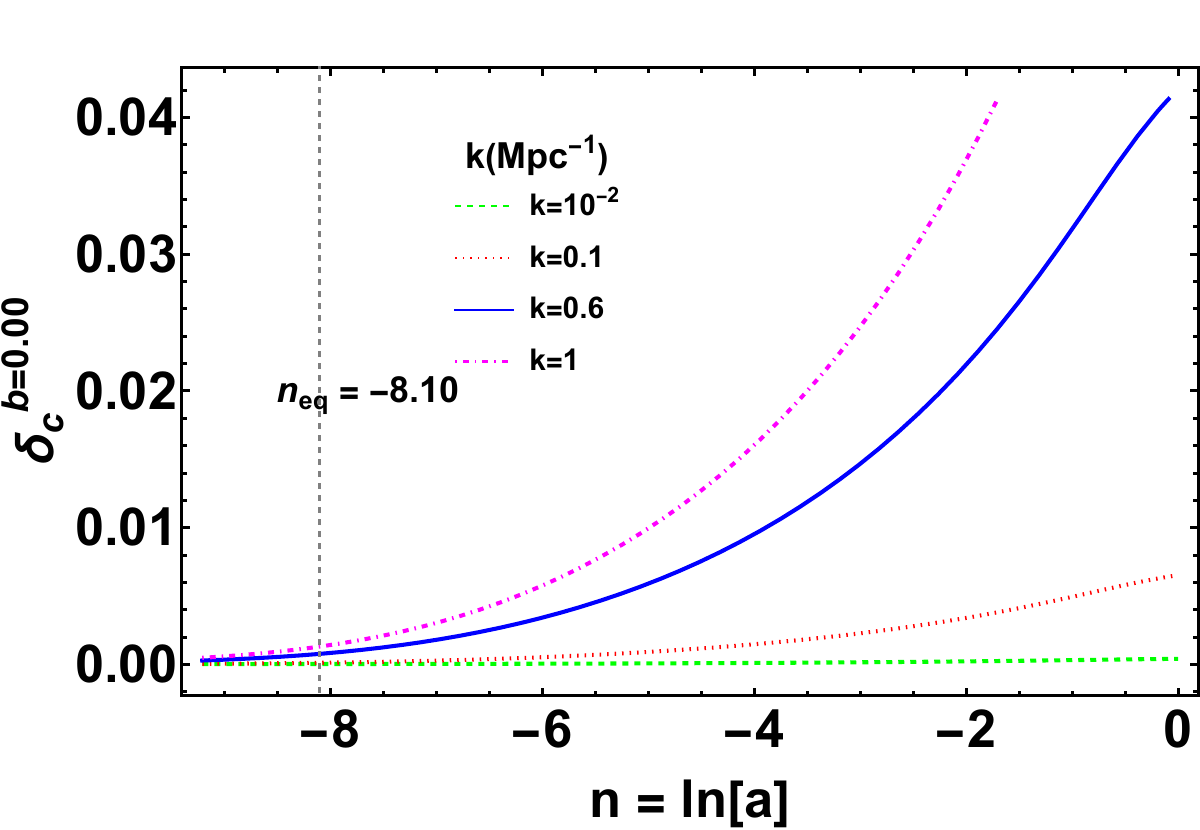} &
\includegraphics[width=0.5\linewidth]{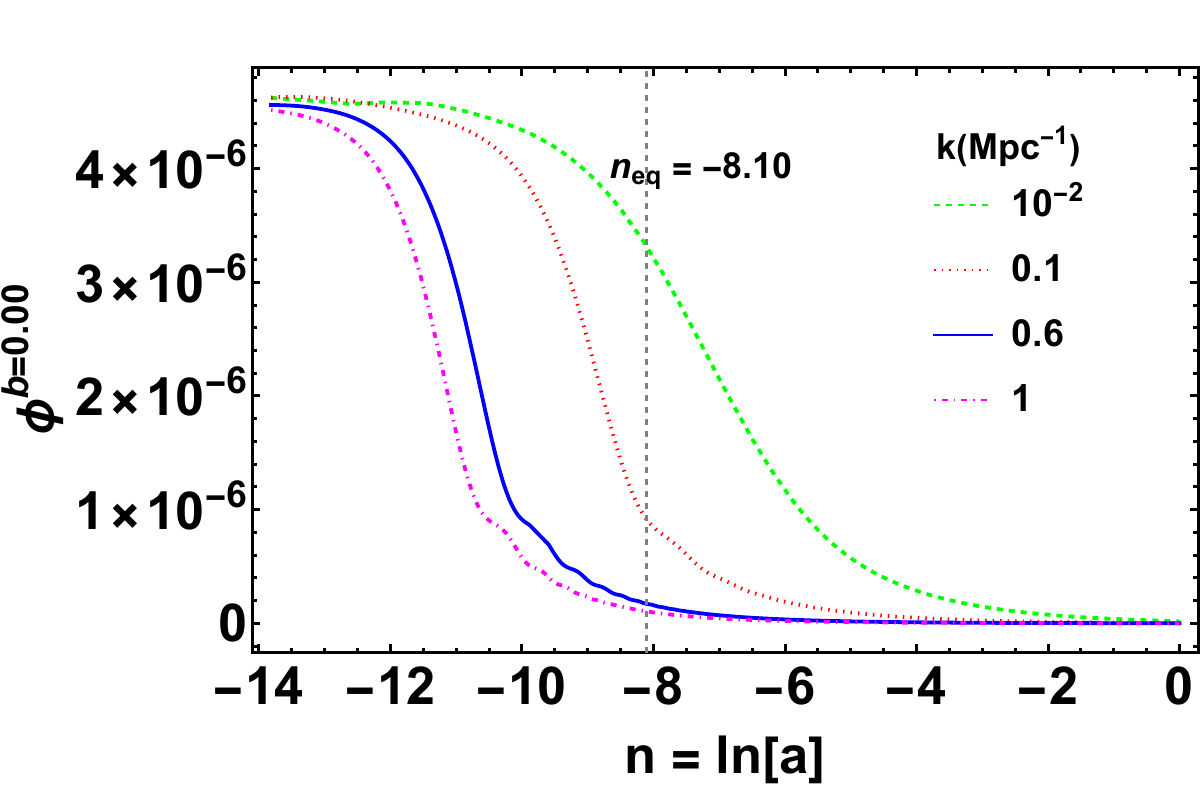}
\end{tabular}
\vspace{-0.5cm}
\caption{ The evolution of dark matter density contrast $\delta_c$ (left panel) and curvature potential $\phi$ (right panel) as a function of the e-folding number $n = \ln a$ for different wavenumbers: $k = 10^{-2}$ Mpc$^{-1}$ (dashed), $0.1$ Mpc$^{-1}$ (dotted), $0.6$ Mpc$^{-1}$ (solid), and $1$ Mpc$^{-1}$ (dot-dashed) for $b = 0$.  a) The dark matter density contrasts for corresponding modes b) The curvature potentials for different modes.} \label{fig-1}
\vspace{1cm}
\end{figure*}

We first consider the case where the speed of light remains constant. 
Figure~\ref{fig-1} illustrates the evolution of the cold dark matter density contrast, $\delta_c$ (left panel), and the curvature potential, $\phi$ (right panel), as a function of the e-folding number $n = \ln[a]$ for various wavenumbers $k$ in the standard cosmological model ($b=0$). The vertical dashed line at $n_{eq} = -8.10$ marks the epoch of matter-radiation equality, $a_{eq}$.
The left panel of Figure~\ref{fig-1} displays the evolution of the CDM density contrast, $\delta_c$, for different wavenumbers: $k=10^{-2} \text{ Mpc}^{-1}$ (dashed green), $k=0.1 \text{ Mpc}^{-1}$ (dotted red), $k=0.6 \text{ Mpc}^{-1}$ (solid blue), and $k=1 \text{ Mpc}^{-1}$ (dot-dashed magenta). For modes entering the horizon during the radiation-dominated era (e.g., $k=0.6 \text{ Mpc}^{-1}$ and $k=1 \text{ Mpc}^{-1}$), $\delta_c$ experiences a period of suppressed growth. This is due to the dominant radiation pressure preventing the gravitational collapse of overdensities until after matter-radiation equality. Following $n_{eq}$, DM perturbations begin to grow linearly with the scale factor, as radiation pressure becomes negligible and DM can freely cluster. For modes that are still super-horizon at matter-radiation equality (e.g., $k=10^{-2} \text{ Mpc}^{-1}$ and $k=0.1 \text{ Mpc}^{-1}$), $\delta_c$ exhibits a nearly constant amplitude until they enter the horizon. After horizon entry and post-$n_{eq}$, these modes also begin to grow linearly. The overall growth rate is faster for larger scales (smaller $k$) as they are less affected by pressure support on smaller scales. All modes show an accelerated growth after matter-radiation equality ($n > n_{eq}$), consistent with the standard understanding of structure formation where CDM density perturbations can grow unimpeded once matter dominates the energy density of the universe. The amplitude of $\delta_c$ at later times is larger for smaller $k$ values, indicating that larger scales grow to become more pronounced over time.

The right panel of Figure~\ref{fig-1} shows the evolution of the curvature potential, $\phi$, for the same set of wavenumbers. During the radiation-dominated epoch, the curvature potential $\phi$ remains nearly constant for modes that are super-horizon. As modes enter the horizon, $\phi$ begins to decay. This decay is a well-known feature in the radiation-dominated era, where the pressure of radiation works against gravitational collapse, causing potentials to decay. Larger $k$ modes (e.g., $k=0.6 \text{ Mpc}^{-1}$ and $k=1 \text{ Mpc}^{-1}$) enter the horizon earlier and thus their potential starts decaying earlier and more significantly.
After matter-radiation equality ($n > n_{eq}$), the curvature potential tends to re-stabilize and become approximately constant again, especially for modes that are super-horizon or have recently entered the horizon. This constant potential is a characteristic of the matter-dominated era, indicating that perturbations grow in a self-similar manner, and the gravitational potential wells deepen as structures form. For smaller scales (larger $k$), $\phi$ has already decayed significantly before $n_{eq}$ and remains at a very small value, approaching zero. The decay of $\phi$ during radiation domination and its constancy during matter domination are crucial for understanding the Integrated Sachs-Wolfe (ISW) effect observed in the CMB. The change in potential across the photon's path contributes to temperature anisotropies. The varying amplitudes of $\phi$ with $k$ are consistent with the scale-dependence of gravitational potential evolution. These figures collectively confirm the standard behavior of cosmological perturbations in a universe without a varying speed of light, providing a baseline for comparison with alternative models.

\subsubsection{For a varying speed of light (\textit{i.e.},  $b \neq 0$)}

\begin{figure*}
\centering
\vspace{1cm}
\begin{tabular}{cc}
\includegraphics[width=0.5\linewidth]{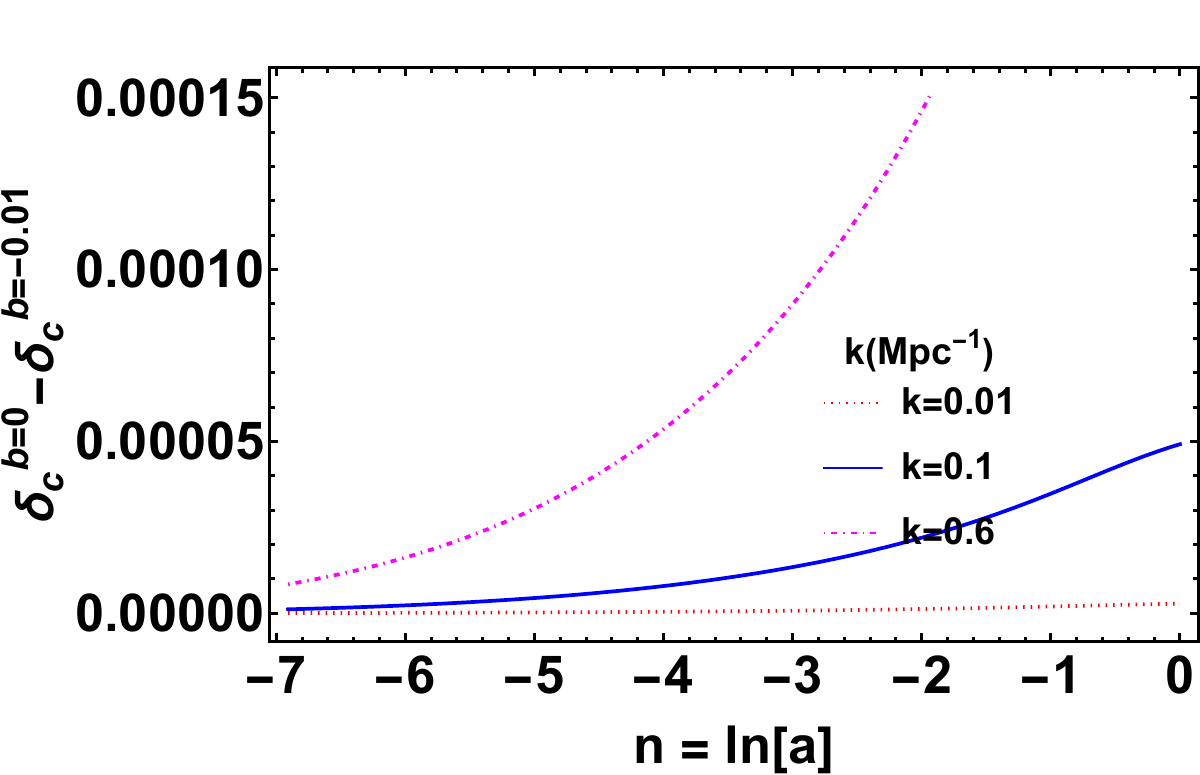} &
\includegraphics[width=0.5\linewidth]{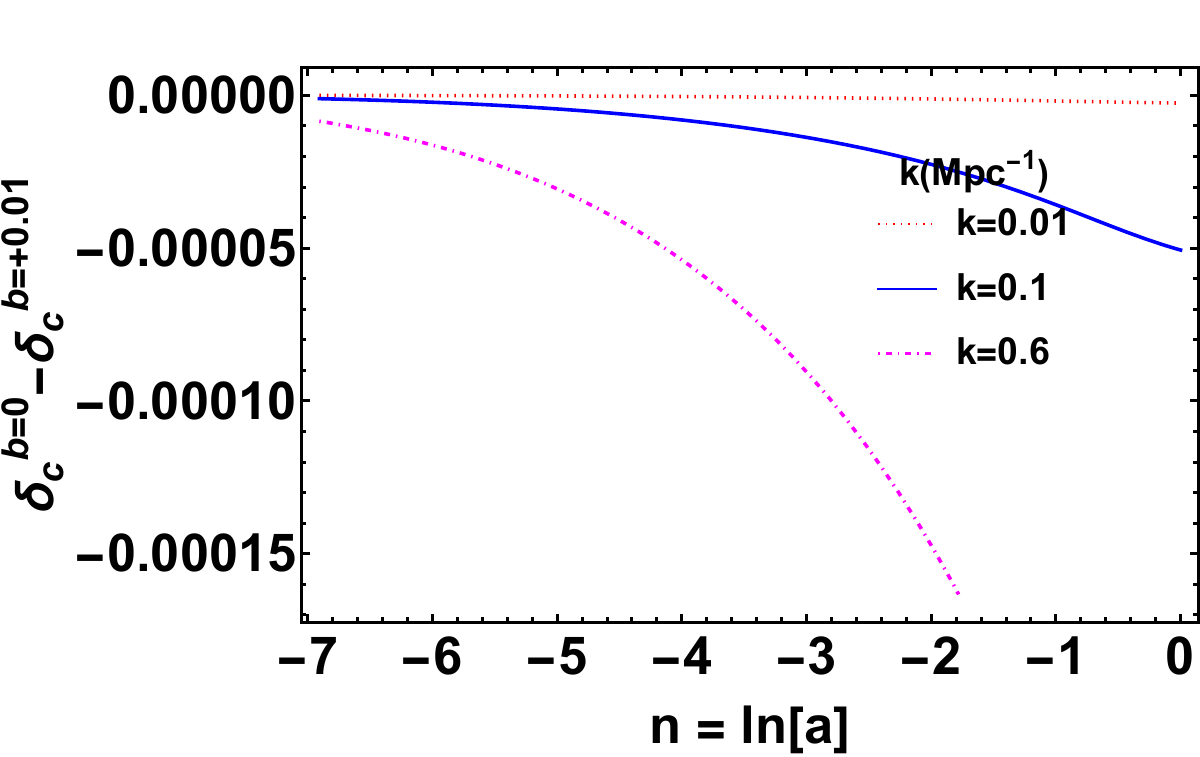}
\end{tabular}
\vspace{-0.5cm}
\caption{ The figure shows the difference in the DM density contrast $\delta_c$ between the varying speed of light cases ($b = \pm 0.01$) and the standard case ($b = 0$).  a) Comparison with $b = -0.01$ (faster past speed of light) case.  b) Comparison with $b = +0.01$ (slower past speed of light) case. } \label{fig-2}
\vspace{1cm}
\end{figure*}

Figure~\ref{fig-2} presents the difference in the CDM density contrast, $\delta_c$, between meVSL models and the SMC ($b=0$). The left panel illustrates the comparison with the $b=-0.01$ case, representing a faster past speed of light, while the right panel shows the comparison with the $b=+0.01$ case, indicating a slower past speed of light. 

The left panel displays the difference $\delta_c^{b=0} - \delta_c^{b=-0.01}$ as a function of $n$ for different wavenumbers $k$. For all investigated wavenumbers ($k=0.01 \text{ Mpc}^{-1}$, $k=0.1 \text{ Mpc}^{-1}$, $k=0.6 \text{ Mpc}^{-1}$), the difference $\delta_c^{b=0} - \delta_c^{b=-0.01}$ is positive and generally increases with $n$. This indicates that in the $b=-0.01$ model, the growth of DM density perturbations ($\delta_c^{b=-0.01}$) is suppressed compared to the standard $b=0$ model ($\delta_c^{b=0}$). The effect is most pronounced for larger scales (smaller $k$). The difference for $k=0.01 \text{ Mpc}^{-1}$ (dotted magenta) is significantly larger than for $k=0.1 \text{ Mpc}^{-1}$ (solid blue) and $k=0.6 \text{ Mpc}^{-1}$ (dotted red). This suggests that the impact of a faster past speed of light on DM clustering is more substantial on larger sub-horizon scales. A faster speed of light in the early universe (corresponding to $b=-0.01$) implies that causal horizons were larger. This can lead to a more efficient smearing out of initial density fluctuations on large scales, or it might alter the gravitational interaction strength and effective sound horizon, thus inhibiting the growth of density perturbations compared to the standard model. Furthermore, the term $+\frac{b}{4} \theta_c$ in the $\theta_c'$ equation becomes $-\frac{0.01}{4}V_c$, potentially leading to a larger decay of velocity perturbations or a more complex interplay with the gravitational potential terms that ultimately results in reduced density growth.

The right panel shows the difference $\delta_c^{b=0} - \delta_c^{b=+0.01}$ as a function of $n$. In contrast to the $b=-0.01$ case, the difference $\delta_c^{b=0} - \delta_c^{b=+0.01}$ is negative for most of the evolution, and its absolute value tends to increase with $n$. This implies that in the $b=+0.01$ model, the growth of dark matter density perturbations ($\delta_c^{b=+0.01}$) is enhanced relative to the standard $b=0$ model ($\delta_c^{b=0}$). Similar to the previous case, the effect is more prominent on larger scales (smaller $k$). The curve for $k=0.01 \text{ Mpc}^{-1}$ (dotted magenta) shows the largest negative difference, indicating the most significant enhancement of $\delta_c$ in the $b=+0.01$ model on these scales. The curves for $k=0.1 \text{ Mpc}^{-1}$ (solid blue) and $k=0.6 \text{ Mpc}^{-1}$ (dotted red) show smaller negative differences. A slower speed of light in the early universe (corresponding to $b=+0.01$) means that causal horizons were smaller. This could potentially localize gravitational interactions more effectively or alter the effective Jeans length, thereby fostering the growth of dark matter perturbations. The term $+\frac{b}{4}V_c$ in the $V_c'$ equation becomes $+\frac{0.01}{4}V_c$, which would tend to counteract the decay of velocity perturbations (if $V_c$ is positive), contributing to a more robust growth of density perturbations compared to the standard model. The enhanced clustering on large scales suggests that the altered causality or gravitational interactions in the $b=+0.01$ model play a significant role in promoting structure formation.

\begin{figure*}
\centering
\vspace{1cm}
\begin{tabular}{cc}
\includegraphics[width=0.5\linewidth]{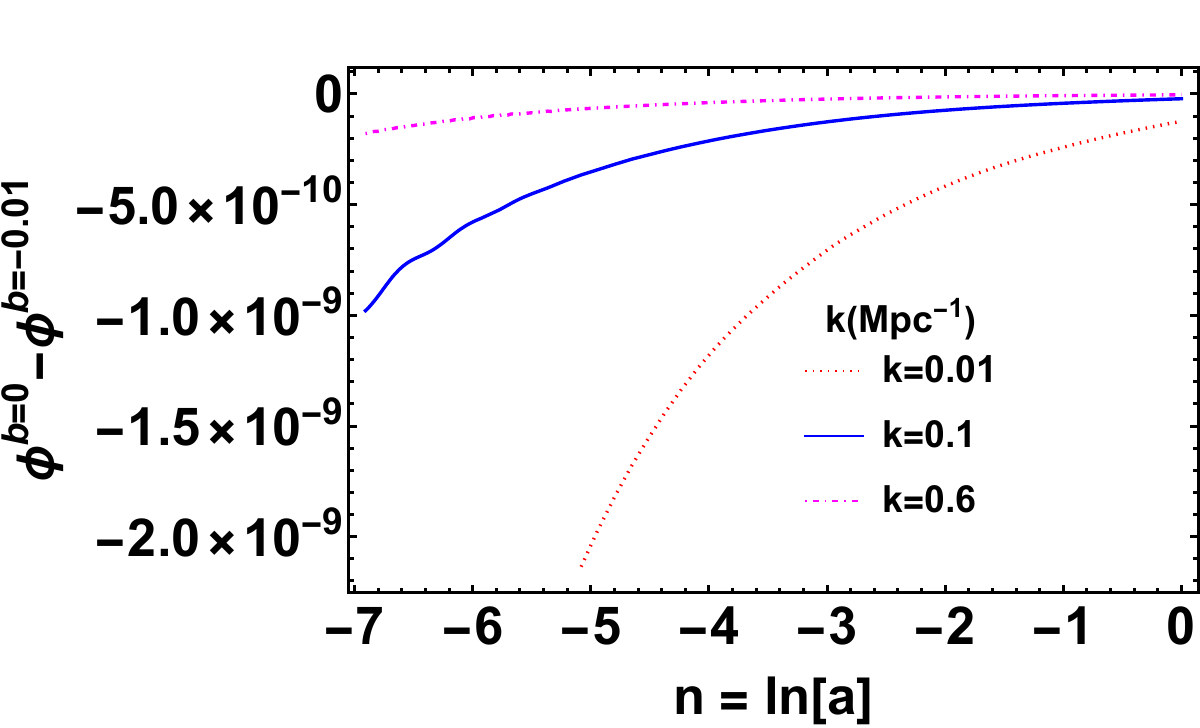} &
\includegraphics[width=0.5\linewidth]{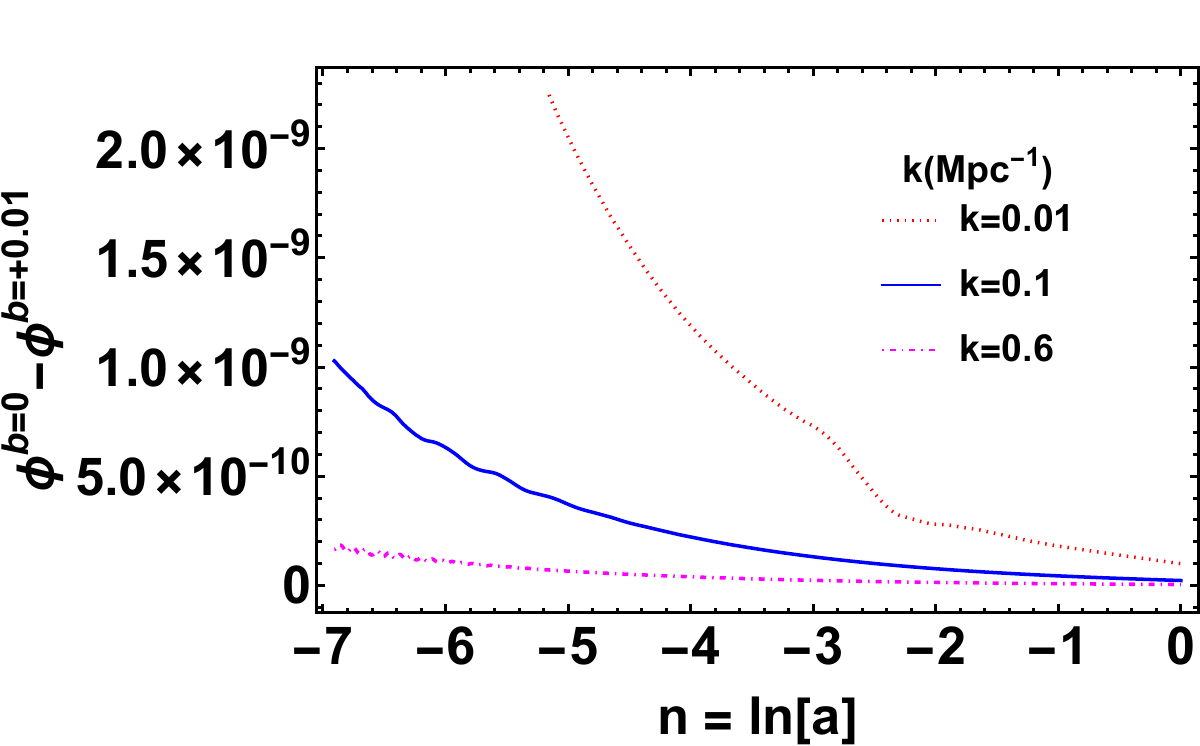}
\end{tabular}
\vspace{-0.5cm}
\caption{ Difference in the curvature potential $\phi$ between the varying speed of light cases ($b = \pm 0.01$) and the standard case ($b = 0$).  a) The case of $b = -0.01$. b) $b = +0.01$ case.} \label{fig-3}
\vspace{1cm}
\end{figure*}

In Figure~\ref{fig-3}, we illustrates the difference in the curvature potential, $\phi$, between meVSL models and the SMC ($b=0$).  The left panel displays the difference $\phi^{b=0} - \phi^{b=-0.01}$ as a function of $n$ for different wavenumbers ($k=0.01 \text{ Mpc}^{-1}$ (dotted), $k=0.1 \text{ Mpc}^{-1}$ (solid), $k=0.6 \text{ Mpc}^{-1}$ (dot-dashed)). The difference $\phi^{b=0} - \phi^{b=-0.01}$ is predominantly negative, with its absolute value generally decreasing as $n$ increases towards the present epoch ($n=0$). This indicates that $\phi^{b=-0.01}$ (the curvature potential in the faster past speed of light model) is less negative, i.e., its absolute value is smaller than $\phi^{b=0}$ (the standard model's potential). In other words, the gravitational potential wells in the $b=-0.01$ model are shallower compared to the standard model. The impact is most significant for larger scales (smaller $k$). The curve for $k=0.01 \text{ Mpc}^{-1}$ shows the largest negative difference, implying the most substantial reduction in the depth of the gravitational potential wells on these scales. The effect diminishes for smaller scales, with $k=0.6 \text{ Mpc}^{-1}$ showing the smallest deviation from zero. The early-time oscillations for smaller $k$ modes suggest a more complex initial response to the varying speed of light. Since the gravitational potential is sourced by density perturbations (as seen in the $\phi'$ equation), a shallower potential well in the $b=-0.01$ model is consistent with the suppressed growth of dark matter density perturbations observed in the left panel of Figure~\ref{fig-2}. A faster past speed of light could lead to a more effective spreading out of initial density fluctuations, resulting in less pronounced overdensities and, consequently, shallower gravitational potentials. The larger effect on larger scales is expected, as these scales are more directly influenced by changes in the fundamental gravitational interactions or causal horizon.
The right panel illustrates the difference $\phi^{b=0} - \phi^{b=+0.01}$ as a function of $n$. The difference $\phi^{b=0} - \phi^{b=+0.01}$ is positive throughout the evolution, with its value generally decreasing as $n$ increases. This implies that $\phi^{b=+0.01}$ (the curvature potential in the slower past speed of light model) is more negative,(\textit{i.e.}, its absolute value is larger than $\phi^{b=0}$.) This indicates that the gravitational potential wells in the $b=+0.01$ model are deeper compared to the standard model. As before, the effect is more pronounced on larger scales (smaller $k$). The curve for $k=0.01 \text{ Mpc}^{-1}$ exhibits the largest positive difference, signifying the most significant increase in the depth of gravitational potential wells on these scales. The influence becomes less pronounced for smaller scales, with $k=0.6 \text{ Mpc}^{-1}$ showing the smallest deviation from zero. The deeper gravitational potential wells in the $b=+0.01$ model are consistent with the enhanced growth of dark matter density perturbations shown in the right panel of Figure~\ref{fig-2}. A slower past speed of light might lead to more localized gravitational interactions, allowing overdensities to collapse more efficiently and form deeper potential wells. The positive difference suggests that the absolute magnitude of the potential, which is typically negative, is larger in the $b=+0.01$ model, indicating stronger gravitational clustering. The stronger effect on large scales is again consistent with the notion that these scales are primarily governed by gravitational dynamics and are sensitive to alterations in fundamental constants.

Figure~\ref{fig-3} clearly demonstrates that a varying speed of light impacts the evolution of the curvature potential. A faster past speed of light ($b=-0.01$) leads to shallower gravitational potential wells, while a slower past speed of light ($b=+0.01$) results in deeper potential wells. These findings are directly linked to the altered growth of DM density perturbations in these meVSL models.

\section{Conclusion and Summary}\label{sec:Conc}

In this study, we have investigated the evolution of cosmological perturbations within the framework of the minimally extended Varying Speed of Light (meVSL) model, focusing on the dark matter density contrast ($\delta_c$), its velocity perturbation ($V_c$), and the Newtonian gravitational potential ($\phi$). By employing the tight coupling approximation for the baryon-photon fluid and utilizing a specific ansatz for the time-varying speed of light, $c = c_0 \exp[(b/4)n]$, we derived and analyzed the perturbed Boltzmann equations. Our analysis considered the standard $\Lambda$CDM model ($b=0$) as a baseline and explored two meVSL scenarios: $b=-0.01$ (implying a faster past speed of light) and $b=+0.01$ (implying a slower past speed of light).

Our findings demonstrate that the varying speed of light leads to relatively larger changes in the dynamics of cosmological perturbations, particularly on large scales. Compared to the standard model, the $b=-0.01$ case shows a suppression in the growth of $\delta_c$, indicating that a faster past speed of light hinders the formation of large-scale structures. Conversely, the $b=+0.01$ case exhibits an enhancement in $\delta_c$ growth, suggesting that a slower past speed of light promotes structure formation. These effects are more pronounced on larger scales (smaller $k$), where gravitational dynamics dominate.

The gravitational potential $\phi$ in the $b=-0.01$ model is found to be shallower (smaller absolute value) than in the standard model. This is consistent with the suppressed density growth, as less clustering leads to less deep potential wells. In contrast, the $b=+0.01$ model results in deeper gravitational potential wells (larger absolute value), reflecting the enhanced density growth. The sensitivity of $\phi$ to the value of $b$ is particularly evident on larger scales, which are most susceptible to changes in the gravitational coupling and causal horizon.

The observed deviations from the standard cosmological model are directly attributable to the additional $b$-dependent terms in the Boltzmann equations, particularly in the velocity evolution of dark matter and baryon-photon fluid, and indirectly through their impact on the gravitational potential. A varying speed of light fundamentally alters the causal structure of the early universe, influencing the propagation of perturbations and the efficiency of gravitational clustering. The scale-dependent nature of these effects suggests that observational probes sensitive to large-scale structures, such as galaxy surveys and cosmic microwave background anisotropies (e.g., through the Integrated Sachs-Wolfe effect), could potentially constrain the parameter $b$ and thus test the validity of the meVSL model.

We acknowledge that our current framework utilizes the tight coupling approximation throughout cosmic evolution. While this provides a fundamental understanding of the VSL impact, a more comprehensive treatment incorporating baryon-photon decoupling and independent baryon evolution would refine the quantitative precision in the post-recombination era. However, our results clearly demonstrate the qualitative deviations induced by a varying speed of light even within this simplification, particularly highlighting its influence on early-time perturbations ($n \lesssim -7$).

Given that current observational data exhibit several tensions that might be explained by the meVSL model \cite{Lee:meVSL}, a more precise and consistent study of perturbation models within meVSL could provide valuable insights into the robustness of the standard cosmological paradigm and the potential for new physics beyond it. Further work would involve incorporating more comprehensive physics, such as neutrino perturbations and anisotropic stress, and performing a detailed comparison with current cosmological datasets to fully assess the viability and observational signatures of the meVSL model.

\section*{Acknowledgments}
SL is supported by the National Research Foundation of Korea (NRF), funded both by the Ministry of Science,  (Grant No. RS-2021-NR059413) and by the Ministry of Education (Grant No. NRF-RS202300243411).  We are grateful to the anonymous referees for their constructive and insightful comments, which significantly improved the quality of this manuscript.



\end{document}